# NoSQL Databases

## Massimo Carro

*Politecnico di Milano*

Piazza Leonardo da Vinci 32

20133 Milano

massimo.carro@mail.polimi.it

# 1. Introduzione

L'arrivo di internet negli anni 90 ha permesso a molte aziende di ampliare il proprio bacino di utenza portando online un'enorme quantità di contenuti e servizi; il caso più esemplare che possiamo citare è Amazon, nata solo nel 1994 ma con un fatturato attestato a 61 miliardi di dollari nel 2012[1].

Di conseguenza, la quantità dei dati presenti sulla "rete" è aumentata esponenzialmente dai *20TB del 1994* passando dai *681PB nel 2003* fino ai *31.338PB del 2012*[2], come si può vedere dal grafico in *figura 1*.

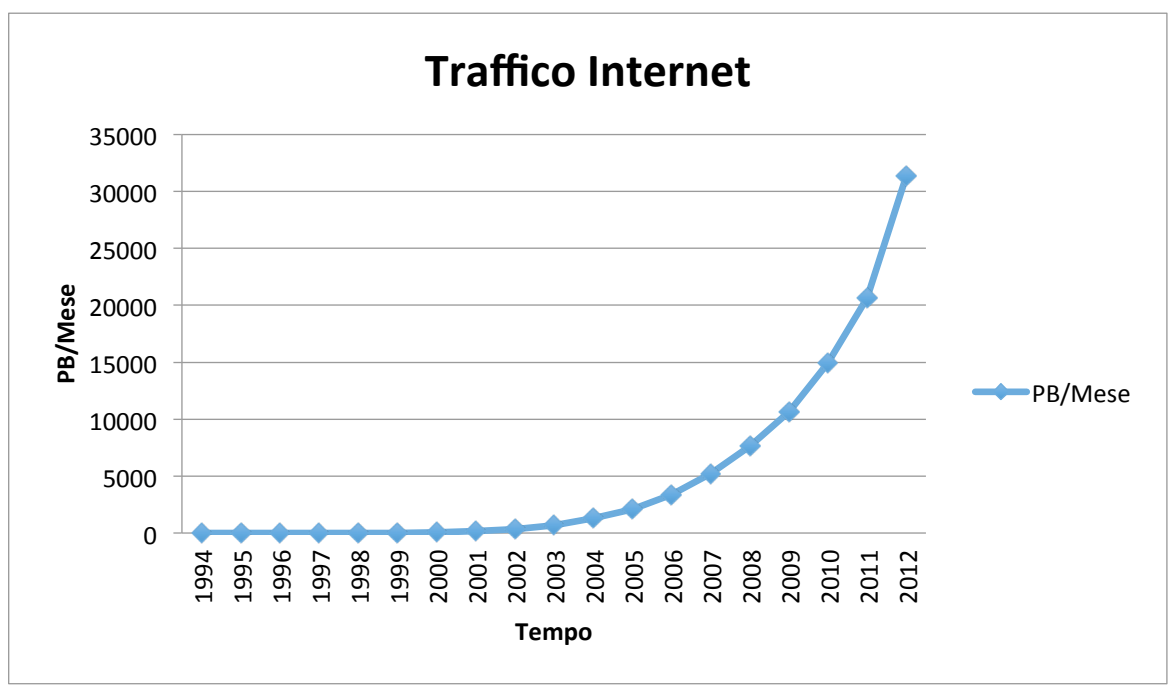

*Figure 1 - Evoluzione del traffico su internet negli ultimi 20 anni.*

Questo ha fatto si che i sistemi informatici debbano elaborare molte più informazioni e quindi maggiore potenza di calcolo, mentre i dati, sempre più numerosi, maggiore possibilità di organizzazione.

Negli ultimi anni con l'avvento del web 2.0, sono aumentate le aziende che operano **esclusivamente** online (si pensi a *Facebook* ed a *FourSquare* ad esempio) fornendo contenuti sempre più numerosi e strutturati; la disponibilità e affidabilità dei servizi unita alla capacità di gestire una grande mole di contenuti e fortemente correlati è diventato un requisito imprescindibile.

Sempre più spesso le aziende on-line e off-line hanno reso più stringenti gli SLA (Service Level Agreement) traducendo a livello contrattuale le aspettative dell'utente finale, ovvero che il servizio sia sempre disponibile, facilmente accessibile (sia come tempi di risposta che come usabilità) e possibilmente attendibile.

La tendenza è diventata quella di garantire più il servizio o il contenuto rispetto della correttezza delle informazioni o alla consistenza dei dati, fatta eccezione dei casi di servizi sotto protocollo sicuro: ad esempio il settore bancario, l'e-governance o alcune fasi di sicurezza dei dati nell'e-commerce.

I fautori di questo nuovo approccio sono stati *Google* e *Amazon* con i loro sistemi proprietari (rispettivamente *BigTable* e *Dynamo*) ma da subito, soprattutto in ambito "*social*", si sono sviluppati sistemi di gestione equiparabili e diversi dai DBMS Relazionali (da ora: RDBMS).

A questi nuovi sistemi è stato dato il nome di DBMS NoSql (da ora: NoSql) proprio per specificare cosa viene a mancare, sia come garanzia che come rigidità: le relazioni.

L'oggetto di questa analisi sono quindi i database NoSql: ne analizzeremo le caratteristiche, i principi fondanti e le opportunità che offrono rispetto ad un RDBMS.

Analizzeremo quindi la struttura di alcuni NoSql per capire come si differenziano tra di loro in funzione dell'uso che se ne vuole fare.

Forniremo inoltre un dettaglio più pratico e operativo di alcuni NoSql (*MongoDB, Riak, Cassandra, e Neo4j*), mostrando tramite delle tabelle comparative i risultati dei test effettuati.

Le motivazioni personali che mi hanno spinto a questa analisi nascono dalla convinzione che questa metodologia di mantenimento ed uso delle informazioni sia un'alternativa spesso sottovalutata in ambito aziendale, mentre l'utilizzo di questi sistemi hanno portato a casi di successo ormai noti come i già citati *Facebook, Amazon*, ma anche *LinkedIn, Ebay* e in ambito nazionale *Blogo*.

## 1.1. Contributi

Dopo una breve introduzione delle caratteristiche salienti di DBMS Relazionali e NoSQL, ci focalizzeremo su questi ultimi per andare nel dettaglio ed analizzare come vengono garantite la *disponibilità delle informazioni*, la *gestione della concorrenza*, la *consistenza dei dati* e la *scalabilità*.

Quello che vogliamo approfondire sono le caratteristiche dei database NoSQL; lo faremo mostrando le differenze che le varie classificazioni (Capitolo 5) hanno tra di loro utilizzando specifici DBMS quali *MongoDB, Riak, Cassandra* e *Neo4i*, uno per ogni categoria[a].

Ci si focalizzerà poi sulla realizzazione di un cluster per ognuno di essi, grazie al quale potremo vedere come i dati possono essere archiviati e/o recuperati e come essere distribuiti nel cluster passando dal concetto di replica a quello di sharding.

Tramite un inserimento massivo vedremo infine i tempi necessari ad inserire le informazioni per ognuno di essi e il carico che le macchine devono sopportare.

Alla fine le conclusioni evidenzieranno pregi e difetti di ogni tipologia esaminata, nonché una breve considerazione sugli ambiti di competenza per gli stessi.

---

[a] *La scelta è ricaduta su questi specifici NoSQL perché ora molto in voga e open-source.*

# 2. RDBMS vs. NoSql

La rivoluzione di internet e l'era del web 2.0 hanno profondamente trasformato l'approccio alla gestione delle informazioni, incentivando la ricerca di nuove metodologie e strumenti.

Questo perché l'importanza dei dati e il volume sempre crescente degli stessi hanno fatto si che l'attenzione si focalizzasse su cosa effettivamente fosse più importante per un servizio: dalla consistenza dei dati alla disponibilità delle informazioni.

Per fare un esempio pratico si pensi alla classifica dei libri più venduti su un sito di e-commerce (Amazon.com, o per citare un caso italiano, Ibs.it): dal punto di vista dell'utente è preferibile accedere immediatamente alla pagina della classifica (anche se non coerente in ogni momento) piuttosto che dovere aspettare un tempo imprecisato per avere l'informazione consistente o aggiornata al secondo.

E' chiaro che l'interesse della libreria online è garantire la disponibilità dell'informazione anche a costo di avere un'informazione approssimata e non del tutto attendibile: Amazon dichiara questa scelta strategica al fine di ottenere 99.9 percentile di disponibilità del proprio servizio.

In casi come questo, la consistenza e le relazioni fornite dai RDBMS possono diventare un collo di bottiglia per sistemi che prediligono, e che possono permettersi di svincolarsi dall'affidabilità delle informazioni.

## 2.1 Proprietà ACID e BASE

Se la caratteristica principale dei sistemi relazionali è quella di avere a disposizione i dati sempre consistenti, i database NoSql garantiscono altissimi livelli di disponibilità degli stessi nonchè della loro velocità di recupero a scapito della consistenza dell'informazione in ogni circostanza.

Per i database relazionali di parla di proprietà *acide* (ACID) mentre per i database NoSql si passa a parlare di proprietà *base* (BASE).

### 2.1.1. ACID

Le proprietà *acide* indicano le caratteristiche che devono avere le transazioni per essere tali.

- (**A**)tomicity – Atomicità

    L'esecuzione della transazione deve essere o totale o nulla: quindi o si riesce ad applicare tutto quello che la transizione vuole che venga cambiato o la transizione stessa viene disfata; nascono i concetti di *commit* e *rollback.*

- (**C**)onsistency - Coerenza

    Quando inizia la transazione il sistema si deve trovare in uno stato coerente e anche quando termina: chiaramente dipendentemente dal livello di isolamento adottato; vengono quindi garantiti i vincoli di integrità dei dati.

- (**I**)solation - Isolamento

    Ogni transazione è indipendente dalle altre. Il suo fallimento non deve provocare danni ad altre transazioni.

- (**D**)urability - Persistenza

Al *commit* della transazione, le modifiche apportate dalla stessa non dovranno essere perse in nessun modo.

## 2.1.2. BASE – Teorema di CAP

Le proprietà BASE sono state introdotte da *Eric Browers*, autore anche del <u>Teorema di CAP</u>. Questo teorema, date le seguenti proprietà:

- **(C)**onsistency

  Tutti i client vedono sempre gli stessi dati anche se si verificano update contemporanee.

- **(A)**vailability

  Tutti i client riescono ad accedere ai dati, anche se a versioni differenti.

- **(P)**artition Tolerance

  La base di dati può essere suddivisa su più macchine.

afferma che:

*"coppie delle proprietà sopra elencate possono essere garantite a scapito della terza; ovvero **non è possibile avere un sistema consistente sempre disponibile i cui dati siano distribuiti su più macchine"**.*

Graficamente quello che afferma il teorema di CAP è così rappresentabile (figura 2); come si può notare in nessun caso le proprietà vengono tutte garantite[10].

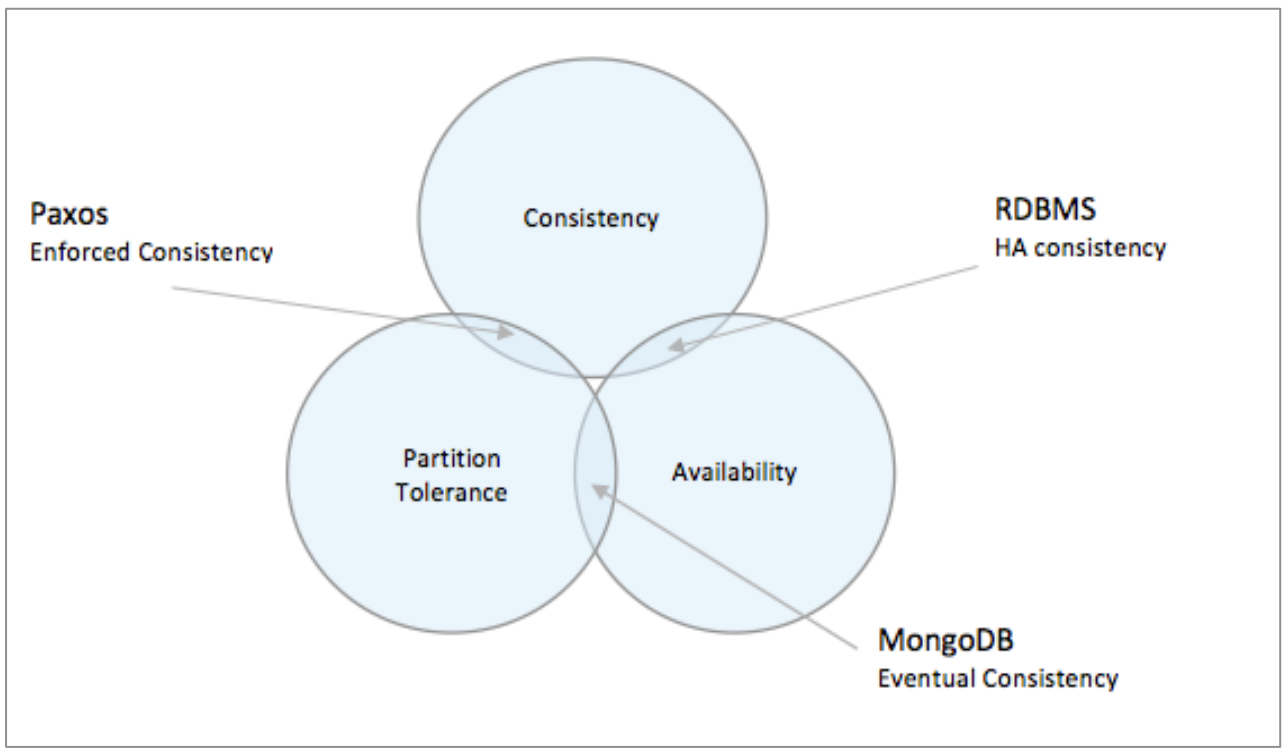

*Figure 2 – Rappresentazione grafica del teorema di CAP ed esempi di adattamento di alcuni strumenti.*

Le proprietà BASE rinunciano alla consistenza per garantire disponibilità dell'informazione e scalabilità.

L'acronimo indica le caratteristiche che un sistema deve garantire sottostando al teorema di CAP.

- **(B)**asically **(A)**vailable

    Il sistema deve garantire la disponibilità delle informazioni.

- **(S)**oft State

    Il sistema può cambiare lo stato nel tempo anche se non sono presenti letture o scritture.

- **(E)**ventual consistency

    In sistema può diventare consistente nel tempo, anche senza ricevere scritture o quant'altro, grazie a dei sistemi di recupero della consistenza.

Fondamentalmente possiamo riassumere le tre proprietà precedenti dicendo che:

*quando la disponibilità dell'informazione è prioritaria, è possibile lasciare che l'applicativo scriva su un nodo quello che deve scrivere senza che si riceva un ok della propagazione di queste scritture sugli altri nodi in cui ci si aspetta venga replicato.*

Questo fa si che si raggiungano alti livelli di disponibilità (non ci sono tempi di attesa) a fronte di quella che definiamo *eventual consistency* ovvero la capacità del sistema di saper sistemare le informazioni sui vari nodi, col passare del tempo, in maniera asincrona.

# 2.2 Pro e Contro RDBMS e NoSql

## 2.2.1 Standardizzazione

L'aspetto più evidente quando si confrontano database relazionali e database NoSql è che i primi appartengono ad uno standard de facto: i risultati di richieste effettuate tramite opportune query possono essere visualizzati in ambienti standard.

Per i NoSql invece non si ha una vera e propria standardizzazione, quindi ogni implementazione ha una interfaccia utente propria e quindi da conoscere ed imparare (lo vedremo più avanti quando analizzeremo alcuni prodotti NoSql).

## 2.2.2 Scalabilità

Un database NoSql, d'altra parte, è più versatile a livello di scalabilità orizzontale, cosa che un RDBMS non può garantire; anzi, quest' ultimo per diventare sempre più "grande" (nel senso di contenere sempre più informazioni) ha necessità di scalare verticalmente e quindi di avere a disposizione hardware sempre più performante e costoso.

I database NoSQL ci consentono di usare macchine molto comuni e non troppo costose, potendo suddividere le informazioni su di esse.

# 3. NoSql : I Principi Fondanti

Come abbiamo già accennato, uno dei motivi che ha spinto verso l'uso dei NoSql è il poter affermare che in alcuni progetti l'integrità dei dati non è prioritaria ne tantomeno obbligatoria; questo fa si che altri tabù vengano a cadere:

- Computer e sistemi di memorizzazione predisposti a garantire performance e sicurezza delle informazioni di altissimo livello, possono essere sostituiti con sistemi dotati di caratteristiche di più facile accesso sia in termini di reperibilità del materiale che di costo.

- L'accesso ai dati diventa molto più rapido: non dovendo sempre garantire una consistenza degli stessi, le letture e le scritture delle informazioni possono essere eseguite senza attendere particolari eventi.

Il primo dei due punti necessita un breve approfondimento: la possibilità di usare hardware di più facilemente disponibile in commercio è legato alla integrazione nei database NoSql di primitive di *sharding* o di partizionamento dei dati; questi sistemi permettono di distribuire le informazioni su più nodi (parleremo più avanti di *cluster*) mantenendo un bilanciamento degli stessi nei casi in cui vengano:

- aggiunti dei dati.

- cancellati dei dati.

- aggiunti dei nodi.

- tolti (volontariamente o a causa di system-failure) dei nodi.

Fatta questo cappello introduttivo, passiamo ora ad elencare, corredandole di informazione, le caratteristiche in possesso dei DBMS NoSql.

- Ambiente Multi-Nodo

    Viene creato un ambiente multi-nodo distribuito a cui viene dato comunemente il nome di anello (o *ring* o *cluster*); questo permette di aumentare o ridurre lo stesso aggiungendo o eliminando dei nodi.

- Sharding dei dati

    Tramite algoritmi appropriati i dati vengono distribuiti su vari nodi e anche recuperati all'esigenza (un algoritmo di esempio è l'algoritmo di gossip).

- Replica delle informazioni

    I dati distribuiti sui nodi vengono anche replicati in modo che le informazioni presenti su ogni singolo nodo siano copiate su altri nodi (solitamente 1 o 2).

- Eliminazione della complessità

    I sistemi RDBMS sono noti per necessitare di infrastrutture complesse per il mantenimento dei dati e l'allineamento degli stessi. Con i sistemi NoSql questa cosa

cambia; la suddivisione dei carichi tra le macchine del cluster fa si che i server possano essere costruiti con hardware di costo più appetibile.

- Correlazione dei dati

    Ad esclusione di casi ad hoc (come vedremo per *Neo4j*) si cerca di far si che sia l'applicativo a correlare le informazioni se necessario; si cerca di mantenere esclusivamente una correlazione se vantaggiosa (come di vedrà con *Cassandra*).

Inoltre molti sono gli studi sui database NoSql al fine di migliorare:

- Performance

    Vengono continuamente studiati algoritmi atti a migliorare complessivamente le prestazioni dei NoSQL.

- Scalabilità Orizzontale

    L'aggiunta di macchine e la rimozione delle stesse dal cluster, deve avvenire senza troppa difficoltà e soprattutto senza che il sistema debba essere fermato o subire momentanei rallentamenti.

- Performance sulla singola macchina

    Oltre a tener da conto la scalabilità orizzontale, vengono effettuati studi per migliorare le prestazioni anche sulle singole macchine, le quali sono in molti casi artefici di ricerca di informazioni nel *cluster* su richiesta degli applicativi.

## 3.1. Replica e Sharding dei dati

La replica dei dati, presente anche nei sistemi RDBMS, è di grande aiuto ai fini di garantire migliori accessi in lettura ai dati oltre che per consolidare la robustezza del sistema complessivo e renderlo, se i dati vengono mantenuti in diversi data center, anche immune da eventi catastrofici quali ad esempio terremoti, uragani, ecc...

Allo stesso tempo avere repliche degli stessi dati su più macchine, fa si che le letture su di essi vengano distribuite sulle varie copie aumentando le performance e rendendo il sistema più robusto, quindi affidabile: se un nodo diventa non disponibile, sappiamo di poter recuperare comunque i dati da una o più macchine che mantengono una replica degli stessi.

Questa implementazione ha ovviamente anche degli svantaggi il primo dei quali è associato alle scritture dei dati: dovendo distribuire i dati su tutti i nodi che mantengono le repliche oltre a quello su cui i dati vengono primariamente salvati, il volume delle scritture aumenta; ciò rende il sistema in alcuni frangenti non coerente a causa degli aumenti dei tempi di latenza per far si che i dati vengano riallineati.

Queste scritture possono essere eseguite sia in modo *sincrono* che *asincrono* e la scelta del metodo ricade su cosa si desidera privilegiare dal punto di vista applicativo.

Ricordando il teorema di CAP:

- Consistenza dei dati (**C**onsistency)

    Dopo ogni operazione il sistema si trova in uno stato consistente: ovvero, per un sistema NoSql, dopo che un'operazione di scrittura è stata effettuata, le letture che seguiranno andranno a leggere correttamente il nuovo dato.

- Disponibilità delle informazioni (**A**vailability)

  Un sistema è costruito in maniera tale che, se un nodo dovesse venire a mancare per qualunque motivo, questa situazione non blocca il sistema.

- **P**artition tolerance

  Il sistema continua a operare anche in seguito ad un partizionamento della rete, inteso come caso in cui due isole di partizioni non riescano più a colloquiare fra di loro.

Correlato alla replica è lo sharding, ovvero il partizionamento dei dati, o se si preferisce la suddivisione in blocchi degli stessi sui vari nodi appartenenti al sistema.

Con lo sharding è possibile suddividere e distribuire i dati su un numero variabile di macchine tali che possano essere aggiunte o tolte a piacere a seconda del carico che le applicazioni portano alla struttura DBMS.

Lo sharding può essere effettuato con vari meccanismi, noi approfondiremo in seguito il *consistent hashing* che permette di suddividere i dati usando una funzione di hash ad hoc sulla chiave primaria.

Lo sharding ha lo svantaggio di rendere complesse le operazioni di join fra i dati, motivo per cui in alcuni NoSql questo tipo di operazione non è effettuabile ma sarà l'applicativo eventualmente a sopperirvi.

La replica più lo sharding creano un meccanismo che riduce sia i costi che i casi di failure del sistema, questi ultimi causati da macchine che si rendono improvvisamente indisponibili ma che non hanno più dati che altre macchine del cluster non possano avere.

# 4. NoSql: Approfondimenti

In questo capitolo approfondiamo i meccanismi usati dai database NoSql al fine di garantire le proprietà alla base di un sistema di gestione dei dati.

## 4.1. Concorrenza

I database NoSql, invece di implementare sistemi di locking, usano altri sistemi per la gestione della concorrenza che permettono di avere meno sicurezza dal punto di vista della correttezza assoluta dei dati ma una maggiore efficienza a livello di prestazioni e tempi di risposta dato che vengono effettuate molte più letture e scritture contemporaneamente.

### 4.1.1. Multi-Version Cuncurrency Control (MVCC)

Il Multi-Version Cuncurrency Control permette di mantenere un controllo della concorrenza basato su continui versionamenti dei dati.

Questi versionamenti sono gestiti tramite tecniche quali i Vector Clocks.

Quando un documento, dopo essere stato letto, necessita di essere modificato, il sistema confronta la versione al tempo della lettura e quella appena prima che la modifica venga apportata: se le due versioni sono diverse significa che un altro processo ha modificato il documento nel mentre.

Da qui si può capire meglio il concetto di multiversionamento: ogniqualvolta un documento viene modificato viene generata una copia (una nuova versione) dello stesso.

Un vantaggio che si può evidenziare è che queste non sono altro che nuove versioni e non modifiche, tutto a vantaggio della velocità; ma d'altro canto, continuare a versionare significa occupare spazio ed è quindi evidente che non potranno essere tenute per sempre tutte le versioni create.

## 4.2. Consistenza

Quando parliamo di consistenza intendiamo il recupero delle informazioni tale che le medesime siano corrette.

Parliamo di *stretta consistenza* nel caso in cui tutte le letture successive ad una scrittura debbano garantire che i dati  siano tali a come la scrittura li ha portati ad essere; quindi o le letture che seguono la scrittura avvengono nello stesso nodo o deve essere utilizzato un protocollo distribuito ad hoc (ad es. *Paxos*) che permetta che si verifichi questa condizione.

Questo tipo di consistenza, ricordando il teorema di CAP, non può essere garantita assieme a disponibilità e fault-tolerance dei dati; questo tipo di consistenza presuppone che si rinunci al partizionamento dei dati o alla disponibilità costante dell'informazione.

Parliamo di *eventual consistency* invece, nel caso in cui le letture possano ritornare dei valori non aggiornati all'ultima scrittura effettuata ma, sicuramente, il sistema prima o poi farò si che questo valore venga correttamente propagato.

Appartengono a questa categoria più tipi di consistenza:

- Read Your Own Write (RYOW) Consistency

    Un client che effettua una scrittura può vedere le letture successive già coerenti indipendentemente dal server su cui la scrittura è stata fatta e le letture successive siano effettuate.

    Questo non vale per gli altri client che eseguono le letture successive alla scrittura del client in oggetto.

- Session Consistency

    Simile alla RYOW con la differenza che le letture successive ad una data scrittura risultano corrette solo all'interno di un certo ambito, ad esempio limitatamente ad un nodo.

- Casual Consistency

    Se un client legge una versione 1 e scrive conseguentemente una nuova versione, tutti i client che leggono la nuova versione proposta dal dato client allora leggevano anche la versione 1.

- Monotonic Read Consistency

    Se un dato client legge un valore per una certa informazione allora, da questa lettura in poi, ogni accesso di qualunque altro client su questa informazione non potrà leggere valori antecedenti.

# 4.3. Versionamento

Nei NoSQL le tipologie di versionamento che vengono implementate sono legate al fatto che le informazioni vengono distribuite su più nodi.

Le varie operazioni di lettura e scrittura possono avvenire su nodi diversi e non può quindi essere garantita una consistenza stretta.

Tra i vari sistemi di versionamento adottati, quelli che andremo a vedere più nel dettaglio sono *Vector Clocks* e *Optimistic Locking*.

## 4.3.1. Vector Clocks

Il Vector Clocks permette di avere un ordinamento di versioni tra le varie operazioni effettuate sui vari nodi.

Si pensi di avere N nodi, ognuno con un suo clock interno che può benissimo essere un contatore che viene incrementato quando serve.

Per ogni operazione effettuata su un nodo, sia essa di lettura o di scrittura, viene incrementato il valore del clock per il nodo stesso. Ogni nodo memorizza al suo interno un vettore (da cui Vector Clocks) che rappresenta lo stato dei vari contatori sui singoli nodi.

Ad ogni operazione eseguita su un nodo, questo vettore, modificato il valore del nodo in cui l'operazione è avvenuta, verrà propagato sui nodi restanti.

Va da se che si generano sequenze che possono garantire di per se la coerenza su tutti i nodi o che purtroppo non possono. In questo ultimo caso sarà un applicativo a dover sbrogliare la situazione e decidere quale nodo definire coerente rispetto ad altri.

Sistemi che usano un sistema di versionamento di questo genere sono *Dynamo* e *Riak*.

## 4.3.2. Optimistic Locking

Nell'Optimistic Locking un unico clock viene salvato per ogni blocco di dati.

Questo tipo di versionamento è stato approfondito dai fautori del progetto Project-Valdemort con la conclusione che in scenari distribuiti dove le macchine vengono aggiunte continuamente ed altre disattivate o rotte improvvisamente, il sistema risulta poco funzionale.

# 4.4. Partizionamento

I sistemi di partizionamento permettono di distribuire le informazioni sul cluster che costituisce il nostro database NoSQL. Ci sono più tipi di partizionamento, di seguito ne elenchiamo alcuni:

- Consistent Hashing
- Memory Cached
- Clustering
- Separating Reads from Writes
- Sharding

Di questi, quelli normalmente più usati rimangono *consistent hashing*  e *sharding* che andiamo a vedere più nel dettaglio.

## 4.4.1. Consistent Hashing

Viene usato per la distribuzione automatica dei dati tra i nodi; garantisce che i dati siano ugualmente distribuiti attraverso il cluster e che l'aggiunta o la rimozione di un nodo in esso possa avvenire automaticamente limitando il numero di spostamenti dei dati.

Il *consistent hashing* usa la stessa funzione di hashing usata per i dati anche per le macchine del cluster; questo permette di calcolare direttamente dal nodo contattato dall'applicazione quale nodo contattare per scrivere o leggere dei dati senza dover passare da un cluster di server dedicati atti a recuperare l'informazione desiderata.

Questa tecnica di partizionamento viene utilizzata da importanti NoSql DB, come *Dynamo, Project Valdemort* (LinkedIn) e  *Riak,* mentre la tecnica che prevede dei server di mapping tra storage e partizioni viene usato da *Google File-System* (GFS) di Google e da *MongoDB*.

Di seguito diamo una veloce dimostrazione di come funziona il consistent hashing.

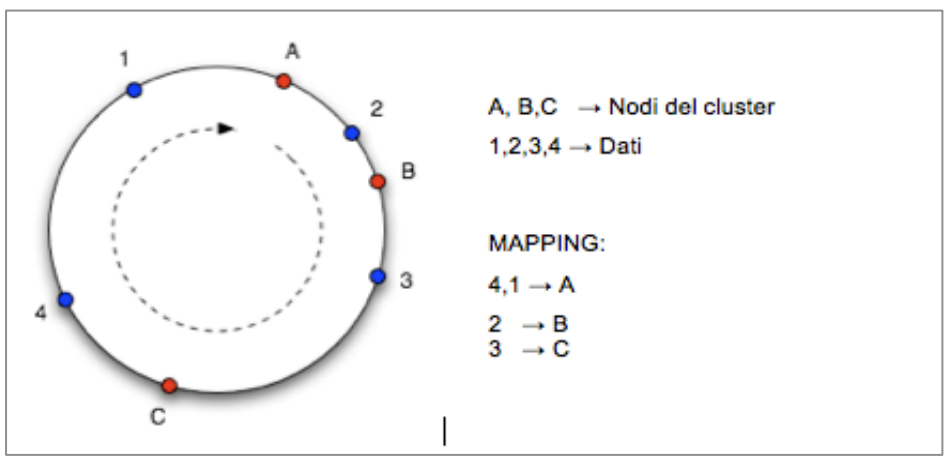

*Figure 4.1 – Consistent Hashing : Distribuzione dell'informazione in un cluster*

Per capire come gli oggetti sono mappati sui nodi basta seguire il giro dell'orologio e si nota che l'oggetto appartenente al nodo precede il nodo stesso.

Supponiamo ora che un nodo abbandoni il cluster per vari motivi, scelta o fail-over che sia.

Avremo che l'oggetto o gli oggetti che precedono il nodo che abbandona lo schema andranno ad appartenere al nodo direttamente successivo ad esso.

Allo stesso modo se un nodo entra nel cluster allora gli oggetti che lo precederanno andranno sotto la sua competenza.

Supponiamo quindi che sia il nodo C ad abbandonare lo schema e a subentrare sia un nodo D, avremo la seguente sitazione.

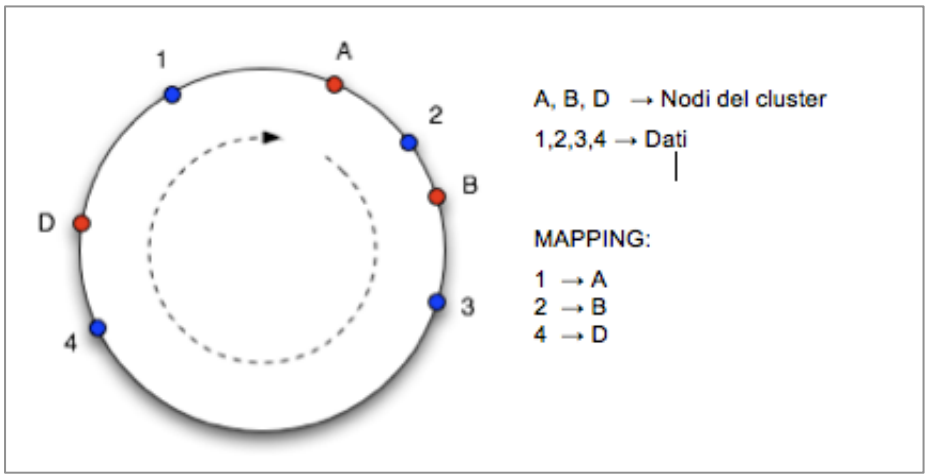

*Figure 4 – Consistent Hashing : Ridistribuzione dell'informazione*

Quello che si nota subito è che al cambiare dei nodi appartenenti al cluster non tutti gli oggetti debbono cambiare nodo di appartenenza ma soltanto alcuni di essi.

Dalla distribuzione che abbiamo dato dei nodi nel cluster rappresentato dalla circonferenza di cui sopra, possiamo vedere che i nodi non sono equamente distribuiti e che quindi può essere che un nodo possa avere più carico di un altro e non essere bilanciati. Per fare fronte a questo inconveniente si è pensato di "inserire" un insieme di *nodi virtuali* per ogni nodo fisico presente nel cluster. Il numero dei nodi virtuali per ogni nodo fisico viene definito da parametri appartenenti alla macchina stessa (*cpu, memoria, dimensione dello storage*) e quindi varia da nodo a nodo.

Il *consistent hashing* garantisce anche un sistema di replica basato su una scelta di progetto dove si fissa un valore K, ovvero un valore numerico che determina quante repliche devono essere mantenute per i dati. Con riferimento alla figura appena sopra, preso ad esempio K=1, i dati 3 e 4 di competenza del nodo D saranno replicati anche sul nodo successivo, ovvero sul nodo A.

## 4.4.2 Sharding

Lo *sharding* è un sistema di distribuzione delle informazioni effettuato sui nodi appartenenti al cluster; queta distribuzione viene dettata da un processo iniziale in cui avviene la definizione della chiave di sharding.

Supponiamo di avere 10 nodi su cui distribuire i dati, possiamo pensare ad una chiave di sharding tale che il modulo della divisione per 10 della chiave primaria (numerica) di ogni dato permetta di decidere dove posizionare il dato.

Questo sistema di distribuzione dei dati viene usato ad esempio su *MongoDB* e necessita di macchine dedicate a questo compito oltre a quelle mantenenti le informazioni.

# 5. Tipologie di database NoSql

Come ogni cosa, tutto ha un inizio, nel nostro ambito ogni modello di database NoSql è stato ispirato da quelli sviluppati ed implementati da Amazon e Google: Amazon con *Dynamo*, Google con *BigTable.*

Le necessità dei due colossi sono state dettate dal bisogno di mantenere *disponibili* e *scalabili* le informazioni in alcuni ambiti che possono andare ad esempio dalla pagina dei libri più venduti a quella della lista dei desideri dei singoli utenti rimanendo nell'ambito e-commerce.

Dynamo[3] appartiene alla categoria dei DBMS NoSQL Key-Value mentre BigTable è a tutti gli effetti un Column Oriented.

Nel mondo open-source si sono poi sviluppate decine, centinaia di DBMS NoSQL (se ne contano ad oggi 150), che si tende a raggruppare in quattro categorie fondamentali:

- Key-Value Distribuiti.
- Document Oriented.
- Column Oriented.
- Graph Oriented.

Questa suddivisione, che ci permetterà analizzare nel dettaglio le caratteristiche per ogni tipologia, racchiude bene o male tutti i database NoSql oggi in circolazione, ma non è mutuamente esclusiva, ovvero alcuni di essi possono avere caratteristiche che appartengono a più di una di queste categorie.

Ad esempio *Riak*, che vedremo essere un Key-Value distribuito, definendo correttamente il tipo dell'informazione da memorizzare, può mantenere anche strutture dati simili ai Document Oriented.

## 5.1. Key-Value Distribuiti

I database NoSql che appartengono a questa categoria permettono di effettuare query unicamente sulle chiavi; queste, essendo per definizione tra loro scorrelate, permetteno di poter costituire un sistema scalabile orizzontalmente.

Un sistema risulta scalabile orizzontalmente quando le coppie <chiave:valore> sono distribuite in maniera appropriata, tramite opportuni algoritmi, sui vari nodi che lo costituiscono.

Il vantaggio di un sistema siffatto è l'altissima concorrenza e quindi la disponibilità delle informazioni, questo a scapito della consistenza dei dati.

Questi database garantiscono replica e fault-tolerance nonché, in molti casi, anche persistenza dei dati su disco e non solo in RAM. Ad esempio, *Riak* ha più sistemi di archiviazione dei dati (detti *backend*) utilizzabili a seconda delle caratteristiche della macchina su cui gira.

Appartengono a questa categoria di NoSql ad esempio:

- Dynamo

- Riak
- Project Valdemort

# 5.2. Document Oriented

La categoria raggruppa quei database che permettono di effettuare delle ricerche, oltre che sulle chiavi, anche sui valori legati ad esse: questo perchè i valori sono rappresentati da strutture in formati pensati proprio a tal fine: *xml* o *json*.

I document oriented usano le proprietà degli oggetti per indicizzarli e per fare ricerche su di essi.

A differenza dei key-value distribuiti, in questo caso il fine ultimo non è il garantire una altissima concorrenza ma il fornire un sistema di interrogazione sui dati performante permettendo la memorizzazione di grossi quantitativi di informazione.

Tra i Document Oriented possiamo trovare:

- CouchDB
- MongoDB
- SimpleDB

# 5.3. Column Oriented

I database che appartengono a questa categoria introducono il concetto di *column-family*.

Una column-family è un'informazione che deve essere presente nello schema (a sua volta quindi necessario) e che serve a raccogliere informazioni di un gruppo; un po' quello che viene fatto per la definizione delle colonne per un database relazionale ma relativo ad un gruppo di colonne, non ad una singola colonna.

Ogni column-family raggruppa un insieme di colonne appartenenti ad uno stessa tipologia di informazione; si pensi ad esempio ad una column-family *"anagrafica"* con colonne costituenti *nome, cognome, età* e *sesso*, o ad una *"indirizzo"* costituita dalle colonne *via, cap, città* e *nazione.*

La column-family deve essere presente nello schema, ma è altresì vero che le colonne che la costituiscono non devono essere definite a priori, ovvero ogni record può avere informazioni diverse per la propria column-family *"anagrafica".*

Supponiamo ad esempio di avere una singola *unità informativa* (quella che in un RDBMS chiamiamo tupla o riga) e che per essa siano definite N column-family nello schema: se una o più informazioni non sono disponibili per *l'unità informativa* in questione, allora le column-family, contenitori di colonne, semplicemente non presenteranno al loro interno le colonne per le quali non si ha la valorizzazione; un po' una situazione *chiave:valore*:

*Se il valore per la chiave esiste allora la colonna è presente ed è nella column-family corretta;*
*se il valore non esiste la colonna semplicemente non esisterà e la column-family ne sarà priva.*

Per avere una visione migliore di quanto esposto si pensi ad un esempio formato da due unità informative (UI), in tal caso due persone, con date informazioni ciascuna.

```
UI : 1
    CF : anagrafica
        nome → Arturo | cognome → Collodi | genere → Maschile | eta → 39
    CF : indirizzo
        via → Pinocchio 1 | cap → 20121 | città → Milano | nazione → Italia
UI : 2
    CF : anagrafica
        nome → Sara | cognome → Piccolo | genere → Femminile
    CF : indirizzo
        via → Alfieri 2 | cap → 20134 | città → Milano
```

*Figure 5.1 – UI : Unità Informativa; CF : Column-Family;*

Con questo esempio riusciamo a vedere la differenza che si può avere tra una archiviazione di tipo RDBMS ed una NoSql: in questi ultimi le colonne che non sono valorizzate non vengono riportate, con un notevole guadagno di memoria rispetto a sistemi relazionali sui quali la creazione di un campo nello schema significa predisposizione al minimo spazio occupabile di memoria o di disco: sui grandi numeri e con poche informazioni a disposizione, il guadagno diventa significativo.

Va tenuto in considerazione che l'aggiunta di una colonna alla column-family, così come l'eliminazione, non comporta una ridefinizione dello schema e quindi è trasparente e immediata; vediamo ad esempio la column-family *anagrafica* dopo dei cambi di colonne presenti al suo interno.

```
CF : anagrafica
UI : 1   nome → Massimo | cognome → Carro | genere → Maschile | eta → 39
UI : 2   nome → Sara | cognome → Piccolo | genere → Femminile | peso ->55
- Aggiunta della chiave hobby e eliminazione del genere nell'UI 1
CF : anagrafica
UI : 1   nome → Massimo | cognome → Carro | hobby → lettura | eta → 39
UI : 2   nome → Sara | cognome → Piccolo | genere → Femminile | peso ->55
```

*Figure 5.2 – Modifica del contenuto delle column-families*

Nei sistemi relazionali una tale aggiunta comporta un cambio nello schema (e quindi un alter table) che su grossi numeri può essere molto onerosa sia a livello di tempo per la pura esecuzione dell'operazione che a livello di memoria allocata se l'informazione aggiunta serve a introdurre informazioni solo per una piccola quantità di righe.

Riassumendo, le caratteristiche principali sono:

- Flessibilità delle colonne appartenenti ad una column-family.
- Partizionamento orizzontale legato alla column-family.
  - Una data column-family non può essere "spezzata" ma ogni record appartenente ad essa può essere su un server diverso.
- Uso di particolari file di log per permettere un più veloce sistema di storage.
- Mancanza di un supporto transazionale.

Alcuni esempi di implementazioni in questa categoria:

- BigTable
- Cassandra
- HyperBase

## 5.4. Graph-Oriented

Le precedenti tipologie focalizzano l'attenzione sulla scalabilità dei dati e sulla flessibilità delle informazioni, hanno però il problema di non poter contenere dati troppo connessi.

Da questa nuova esigenza prende luce l'idea dei NoSQL orientati ai grafi.

Possiamo immaginare questa categorie come una document oriented con la clausola che i documenti sono anche adibiti a rappresentare le relazioni.

Strutture di questo tipo permettono il passaggio tra un nodo ad un altro (graph traversal) e sono quindi molto utilizzati nel campo del Social Network.

Esempi per essi :

- Neo4j
- FlockDb
- Pregel

## 5.5. Tabella di Comparazione

Qui di seguito riassumiamo le caratteristiche principali che contraddistinguono le tipologie viste.

|  | Key-Value | Document-Oriented | Column-Family | Graph-Oriented |
|---|---|---|---|---|
| Schema | NO | NO | SI | NO |
| Correlazione dati | NO | NO | SI | SI |
| Caratteristica principale | Altissima Concorrenza | Query performanti, moli di dati altissime | Singola riga CF non separabile | Presenza delle relazioni tra i dati |

# 6. Confronto tra DBMS NoSql

Introdotte le tipologie NoSQL, andiamo ad analizzare nel dettaglio le differenze che le contraddistingono analizzando più da vicino un DBMS per ognuna di esse:

- **Riak** : Key-Value Distributed.
- **MongoDB** : Document Oriented.
- **Cassandra** : Column Oriented.
- **Neo4J** : Graph Oriented.

E precisamente, costruiremo un ambiente di lavoro costituito da 3 macchine virtuali (formanti un cluster di 3 istanze), per le quali vedremo:

- *Introduzione per ogni NoSQL da analizzare*

  Per ogni DBMS NoSQL che andremo ad osservare, verrà fatta una introduzione al fine di definirne di elencarne le caratteristiche principali.

- *Costruzione del cluster e costo di inizializzazione:* **RAM***, CPU, IO.*

  L'installazione di qualunque DBMS produce dei costi iniziali per far si che esso funzioni anche senza essere utilizzato. In questa sezione vedremo, con la struttura a cluster messa in opera, come i NoSQL in esame sfruttino le risorse del sistema a loro disposizione.

- *Replica, sharding (scalabilità orizzontale) e High Availability (HA) dai dati.*

  Parliamo di _replica_ come la capacità di un sistema di mantenere più copie dei dati con molteplici obiettivi che vanno dall'accesso ai dati di applicativi su più istanze alla copia degli stessi dati per ragioni di sicurezza (fail-over) fino a permettere di avere un sistema di backup che agisce su una di queste copie mentre le altre sono in produzione.

  Con _sharding_ e HA intendiamo invece la possibilità di "spezzare" la base di dati su più macchine tramite opportuni algoritmi, facendo si che le macchine e il numero di istanze possano cambiare più o meno a piacimento (aggiungo un nodo per migliorare la distribuzione dei dai o si rompe una server che mantiene dei nodi) mantenendo il servizio in produzione.

- *Connessione, accesso ai dati, caricamento massivo.*

  Ogni NoSQL ha una propria console di accesso ai dati e dei drivers, per il collegamento dei vari applicativi, scritti su specifici linguaggi di programmazione quali *Python*, *Java*, *Ruby*, *PHP*, ecc…

  In questa sezione vedremo di analizzare cosa la propria console di accesso ai dati dia in senso di accessibilità di informazione e cosa il NoSQL in esame permetta di fare sui dati che mantiene senza dover scomodare applicativo o il programmatore. Inoltre, tramite degli scripts in *Python* effettueremo degli inserimenti massivi di dati analizzandone i risultati.

- *Backup dei dati.*

  Ovvero il meccanismo tramite cui il sistema permette di avere una copia dei dati di sicurezza.

Alla fine di tutte queste analisi, riassumeremo i risultati in una tabella di facile osservazione.

# 6.1. Introduzione ai NoSQL scelti

## 6.1.1. MongoDB

MongoDB è un database Document Oriented con possibilità di *replica* e *sharding* dei dati.

E' dotato di una struttura di replica *Primary/Secondaries* che garantisce, sotto certi limiti e tramite una sistema di elezioni appropriato, il fail-over automatico del *Primary* a vantaggio di un dato *Secondary*. La gestione del *fail-over* è possibile fino ad un numero massimo di 12 (un *Primary* e undici *Secondaries*), superato questo numero il sistema si riduce ad un classico *Master/Slave* ed il *fail-over* è da gestirsi manualmente.

Lo *sharding* permette di suddividere i dati tra più strutture *Primary/Secondaries*, quindi di scalare su più *Primary* i dati che poi verranno eventualmente replicati.

Le macchine che mantengono i dati sono chiamate *mongod*, quelle che si occupano di gestire la distribuzione delle informazioni *mongos.*

Comunque sia, le scritture e le letture consistenti sono sempre eseguite sui *Primary*, mentre le letture eventualmente consistenti anche sui *Secondaries*.

Ogni istanza può contenere più database e le strutture interne ad ogni database sono chiamate *Collection*.

## 6.1.2. Riak

Riak è un database Key-Value distribuito e decentralizzato contraddistinto da una struttura ad anello, detta anche *ring* o *cluster*, che integra già *replica* e *sharding*.

Ogni nodo dell'anello è uguale all'altro ed è lui stesso che, se "contattato" da un applicativo, si fà carico di recuperare le informazioni richieste ovunque esse siano, ritornandole al richiedente.

Questo processo avviene tramite algoritmi di *gossiping* che mantengono aggiornate le informazioni della struttura nel tempo.

Riak permette di definire più strutture dette *bucket*, che possono essere paragonate alle tabelle nella logica SQL, ma non permette di creare più di un database in un singolo *cluster*.

## 6.1.3. Cassandra

Cassandra è un database Column Oriented (detto anche Column Family) semistrutturato, ovvero dotato di uno schema caratterizzato da *databases* e  da *Column Families* (CF), che nella versione che analizziamo assumono anche il nome di tabelle;  ogni CF può avere chiavi differenti al suo interno ma se esistono chiavi uguali per tuple differenti, allora le dette chiavi apparterranno sempre alla stessa Column Family (vedi capitolo 5 : Column-Oriented).

Cassandra integra un meccanismo di *replica* e di *sharding*.

Lo *sharding* viene effettuato per *Column Family*, ovvero se una chiave di una CF è situata su una data macchina, allora lo saranno anche le altre chiavi di quella CF per la data tupla di informazioni. Questa tipo di implementazione deriva da analisi che hanno evidenziato che, quando viene richiesta una informazione appartenente ad una data *Column Family*, solitamente vengono richieste anche le altre informazioni presenti in essa.

Per esempio una *Column Family* "indirizzo" conterrà chiavi quali **via, città, numero civico,** ecc… e se recuperiamo le informazioni di anagrafica di un certo utente, alla voce "indirizzo" serviranno tutte queste voci.

Così facendo, l'accesso alla CF "indirizzo" di una data tupla permette di andare a recuperare tutte le informazioni ad esso inerenti sulla stessa macchina e quindi di velocizzare i tempi di risposta.

In Cassandra ogni nodo di un cluster è uguale agli altri: questo fa sì che su ognuno di essi possano essere effettuate letture e scritture.

Da precisare, però, che all'avvio del cluster alcuni nodi hanno dei compiti aggiuntivi; questi nodi prendono il nome di *seeds* ed è loro compito generare il cluster andando a recuperare tutti gli altri nodi che lo costituiscono. Una volta terminato questa fase, tutti i nodi del cluster tornano ad avere le stesse caratteristiche.

Cassandra ha la possibilità di generare più databases per ogni cluster.

### 6.1.4. Neo4j

Neo4j è un database *Graph Oriented* che permette di mantenere sia i dati che le relazioni presenti tra questi; è simile ad un Document Oriented con la differenza che ogni documento può rappresentare dati o relazioni.[b]

Neo4j presenta una struttura Master/Slave con la possibilità di effettuare le scritture su tutti i nodi del cluster.

Non è dotato di sharding e mantiene un unico grafo per ogni cluster.

### 6.1.5. Tabella Comparativa

Di seguito raccogliamo le informazioni, viste nei precedenti sotto paragrafi, per i 4 DBMS NoSQL sotto osservazione.

|  | MongoDB | Riak | Cassandra | Neo4j |
|---|---|---|---|---|
| Struttura | Primary/Secondary | Cluster | Cluster | Master/Slave |
| Replica | SI | SI | SI | SI |
| Sharding | SI | SI | SI | NO |
| Multi DB | SI | NO | SI | NO |

*Figure 6.1.1 – Tabella comparativa NoSQL*

# 6.2. Architettura per l'analisi

Strutturare un caso reale per analizzare le differenze tra i vari NoSQL non è cosa semplice perché vorrebbe dire gestire molti dati, suddividerli su vari server collocati in aree geografiche differenti al fine di migliorarne le performance; ottenuta questa configurazione, poi, iniziare ad evidenziarne le differenze per mantenimento e accesso delle informazioni.

Non potendo generare un ambiente così strutturato, ci limitiamo alla costruzione di un *cluster* costituito da 3 macchine virtuali realizzate tramite un sistema di virtualizzazione ad hoc, *Oracle VirtualBox*.[19]

Queste macchine virtuali verranno generate tramite *Vagrant*[11] che, grazie ad un semplice file di configurazione (*Vagrantfile*) ed appoggiandosi ad un dato sistema di virtualizzazione, ci permette di creare quante macchine virtuali desidereremo su un unico Personal Computer che le ospita.

---

[b] *In maniera esclusiva, un documento può contenere o dati o relazioni.*

La cosa che rende appetibile questo sistema è che aggiungere o togliere una macchina virtuale è molto semplice e consiste in poche righe dentro il file di configurazione di Vagrant stesso.

In appendice si può trovare il file di configurazione usato per la costruzione di questo ambiente.

> Nota
>
> *Ogni cluster verrà avviato a se stante per problemi di performance della macchina locale: quindi un cluster alla volta.*

Il numero 3 di MV è stato deciso tenendo conto dei limiti che il personal computer ha come RAM, velocità di lettura disco e gestione risorse da distribuire fra queste.

Detto ciò, lo schema di partenza per tutti i NoSql sarà quindi rappresentabile come segue:

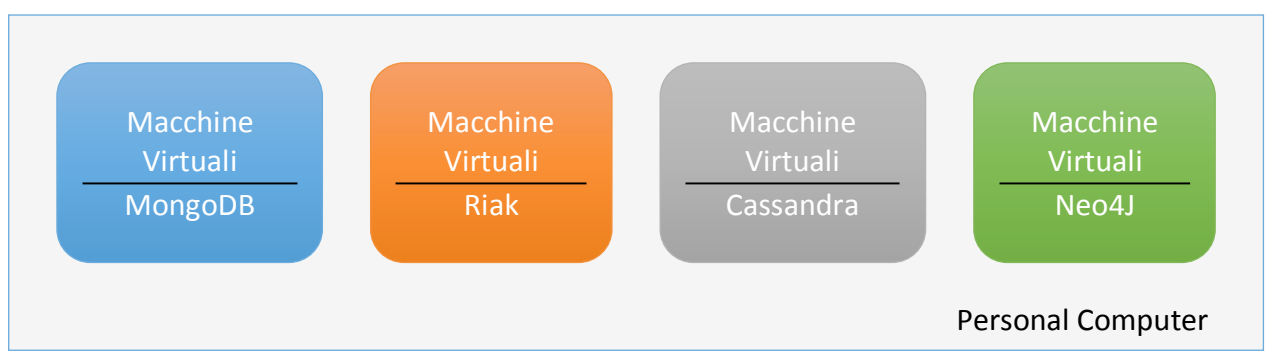

*Figura 6.2.1: Struttura di analisi cluster NoSql*

# 6.3. Costi di inizializzazione – Uso Disco, RAM, CPU

La macchina che ospita le tre macchine virtuali è un MacBook Pro i5 dotato di:

- 500GB hard disk
- 8GB memoria RAM
- 4 CPU

Le macchine virtuali create hanno ognuna:

- 500GB hard disk - condiviso
- 480MB memoria RAM
- 1 CPU

Si ricorda qui che le valutazioni che seguiranno sono effettuate su un ambiente di replica già configurato e funzionante per tutti i DBMS considerati.

## 6.3.1. MongoDB

MongoDB mette a disposizione dei repository ad hoc per l'installazione.

Tramite *yum* è molto facile avere in pochi minuti una macchina stand-alone funzionante, ci basta creare un file */etc/yum.repos.d/mongodb.repo* con il seguente contenuto e lanciare il comando che segue.

```
[mongodb]
name=MongoDB Repository
baseurl=http://downloads-distro.mongodb.org/repo/redhat/os/x86_64/
gpgcheck=0
enabled=1
```

*Figure 6.3.1 – Contenuto del file : /etc/yum.repos.d/mongodb.repo*

```
bash> sudo yum install mongo-10gen mongo-10gen-server
```

*Figure 6.3.2 - Comandi di installazione per MongoDB*

Per una configurazione della replica bisogna lavorare sia a livello si configurazione che a livello di console *mongo* e una volta che questa è configurata *MongoDB* avrà:

- Creato di un database *local* per la gestione della replica (database non replicato ma presente su ognuna delle istanze) e dimensionato gli *oplog* (default 5% del totale disco a disposizione).

- Allocato spazio per i files di *journaling* (default 3 file da 1GB l'uno).

Quindi avendo a disposizione un disco di 500GB, una istanza *mongod* occupa circa 28GB.

Per la RAM invece al boot *mongod* usa circa 300MB e non facendo nessuna attività, se non tenere traccia delle replica per la gestione eventuale del fail-over, fa pochissime letture/scritture da/su disco.

I testi effettuati trattano una versione di *MongoDB 2.4.3*.

Verificare quanto prende di disco per installare MongoDB.

## 6.3.2. Riak

Anche *Riak* come MongoDB può essere installato facilmente tramite *yum.*

I comandi necessari sono i seguenti:

```
bash> sudo yum install http://yum.basho.com/gpg/basho-release-5-1.noarch.rpm
bash> sudo yum install riak
```

*Figure 6.3.3 - Comandi di installazione per Riak*

Anche Riak per inizializzare la replica necessita che sia configurato il cluster dentro il file di configurazione che a livello di console *bash*; inoltre per funzionare con più agio richiede una modifica al numero dei file aperti dall'applicazione che deve essere almeno pari a 4096.

Riak può archiviare le informazioni tramite diverse strutture di backend, quella di default e che noi useremo è *Bitcask.*

Alla partenza, una istanza Riak con replica configurata e archiviazione Bitcask, crea una directory per contenere i dati di circa 5MB e sfrutta una quantità di RAM pari a circa 440M.

Riak è scritto in Erlang e installare Riak prende 45MB disco.

I test sono fatti con *Riak 1.4.2*.

### 6.3.3. Cassandra

Cassandra è un progetto open-source Apache e si basa su JVM.

Le nostre considerazioni si basano su Cassandra 2.0, per la quale versione sono state fatte modifiche sostanziali rispetto le vecchie versioni 1.x.

Scaricando i binari e avendo già JVM installata si può subito far partire una istanza di Cassandra.

```
bash> wget \
"http://www.apache.org/dyn/closer.cgi?path=/cassandra/2.0/apache-
cassandra-2.0.3-bin.tar.gz"
bash> tar zxf apache-cassandra-2.0.3-bin.tar.gz
```

*Figure 6.3.4 - Comandi di recupero ed installazione di Cassandra*

La *replica* viene gestita soltanto modificando i file di configurazione di ogni macchina, una volta partite le macchine la replica è funzionante.

Esistono dei tool che permettono di gestire il cluster di cassandra da interfaccia web, ad esempio *Datastax OpsCenter* o *Helenos*.

Una singola istanza in replica occupa circa 200MB di RAM e 1% di CPU quando parte, pochissimo spazio disco, 400KB.

L'installazione di JVM prende circa 110MB di disco.

I test sono stati fatti con *Cassandra 2.0* e *JVM 1.7*.

### 6.3.4. Neo4j

Neo4j è un progetto di Neo Technology e si basa su JVM.

Per poter funzionare la macchina che lo ospita deve avere, oltre JVM, anche il pacchetto *lsof*.

Si scaricano direttamente i binari e metterlo in funzione è molto veloce.

Richiede che il numero di fie aperti sia almeno 40.000.

```
http://info.neotechnology.com/download/neo4j_enterprise_1.9.5.tar.gz
tar zxf neo4j_version.tar.gz
```

*Figure 6.3.5 - Comandi di recupero ed installazione di Neo4j*

L'esecuzione iniziale occupa 10% di memoria e 2.5% di CPU.

Neo4j ha una interfaccia grafica integrata che risponde sulla porta 7474 di default.

I test sono fatti con *Neo4j 1.9.5 e JVM 1.7*.

### 6.3.5. Comparazioni

I grafici rappresentano l'occupazione di disco, RAM e CPU nel momento in cui una istanza viene avviata e collegata al *cluster*.

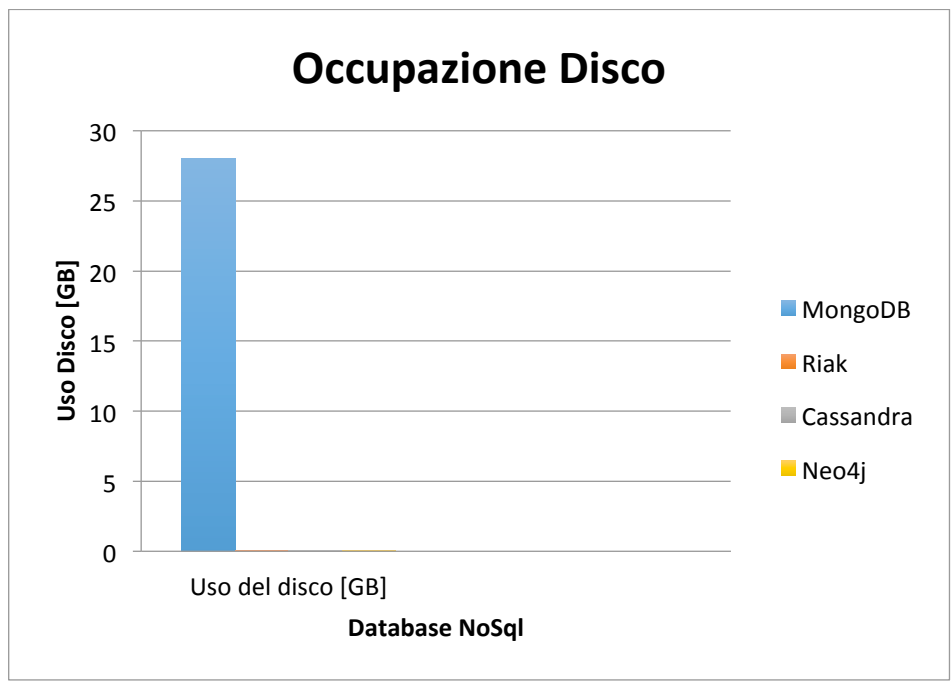

*Figure 6.3.5 - Uso del disco : instanza avviata ma nessuna attività verso di essa. La rappresentazione è molto falsata da MongoDB che occupa qusi 30GB di disco, le altre installazioni sono molto limitate come occupazione. C'è da dire che questa discrepanza è tale anche per come MongoDB parte di default, nulla vieta di abbassare lo spazio disco occupato al boot per esso, ma l'esame viene fatto proprio senza sapere quanto ogni sistema cuba.*

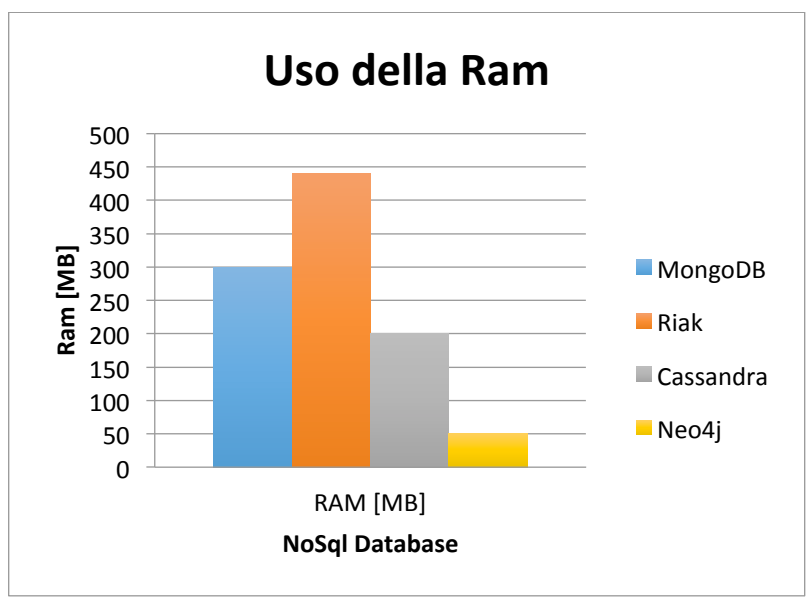

*Figure 6.3.6 - Uso della RAM : instanza avviata ma nessuna attività verso di essa.*

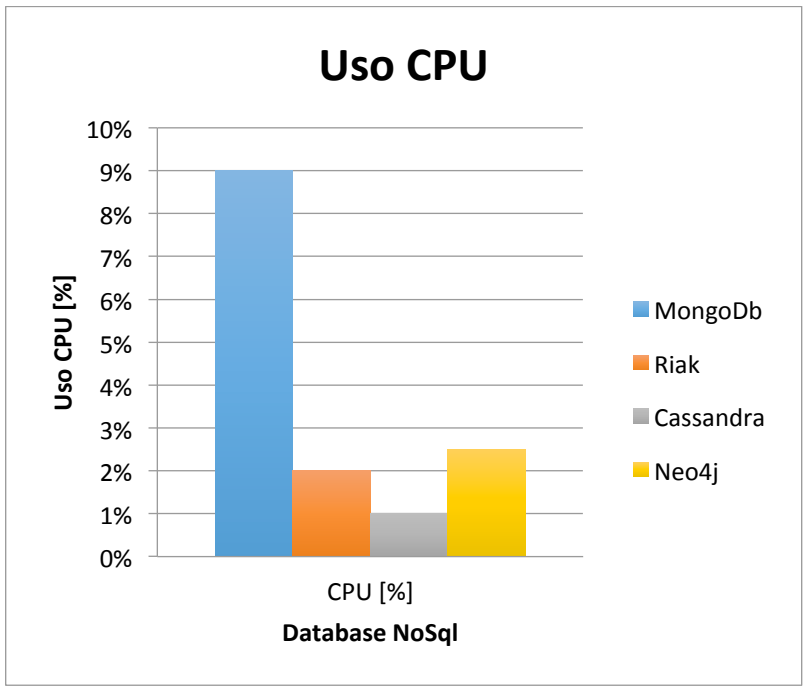

*Figure 4.3.7 - Uso della CPU : istanza in piedi ma nessuna attività verso di essa.*

### 6.3.6 Tabella Comparativa

|  | MongoDB | Riak | Cassandra | Neo4j |
|---|---|---|---|---|
| File aperti minino | Non necessario | 4096 | Non necessario | 40000 |
| Interfaccia grafica | SI | SI | SI* | SI |
| Installazione | yum | yum | binari | binari |
| Pacchetti aggiuntivi | NO | NO | NO | lsof |

*Figure 6.3.8 – Informazioni legate al boot di un istanza in un cluster. *tramite pacchetti esterni al tool.*

# 6.4. Replica, sharding e HA dei dati

Le informazioni che un DBMS raccoglie, possono essere tutte concentrate su una macchina o distribuite tra i nodi del *cluster* costituenti il DBMS.

In quest'ultimo cado si parla di *sharding* dei dati.

Lo *sharding* permette di scalare su più macchine all'aumentare delle informazioni (si parla di *scalabilità orizzontale*). Senza sharding le informazioni continuerebbero a essere inserite su una macchina unica che, dovento garantire scrittura e lettura delle informazioni, dovrà essere sempre più performante (si parla in tal caso di *scalabilità verticale*) col crescere dei dati nel tempo.

Con o senza *sharding*, i dati hanno la necessità di essere replicati su altri nodi soprattutti per garantire che tutte le informazioni presenti su una macchina siano almeno disponibili su un'altra.

Questo permette una gestione appropriata del failover: se una macchina si spegne per qualunque motivo, un'altra ha tutte le carte in tavola per prendere il suo posto.

*Replica* e *sharding* vanno solitamente a braccetto, tanto che, se le informazioni sono distribuite su più nodi (*sharding*), ogni nodo potrà avere i suoi contenuti replicati su altri nodi del cluster (*replica*) ed avere lui stesso delle repliche di altri nodi sempre appartenenti al medesimo cluster.

## 6.4.1. MongoDB

E' un database distribuito ma non decentralizzato, il che significa che è dotato di :

- una macchina principale che chiamiamo *Primary* su cui vengono fatte le scritture e le letture consistenti.

- Altre macchine, *Secondaries*, per letture eventualmente consistenti e su cui vengono replicate le informazioni del *Primary*.

Le configurazioni più usate sono :

1. **Un Primary + N Secondaries**, fino ad un massimo di 12 istanze, incluso il *Primary*; in tal caso la gestione del failover è automatica.

2. **Un Primary + N Secondaries**, fino ad un numero non precisato: questo è il caso a cui degenera il precedente se il numero di secondaries supera le 11 unità; siamo in una configurazione classica *Master/Slave*, il failover è da gestire a mano.

3. **N Mongos + M Primary + S Secondaries (Sharding),** abbiamo in questo caso un numero N di macchine che si occupano di gestire lo sharding dei dati sugli M *Primary* appartenenri alle diverse catene ognuna delle quali del tipo definito al punto 1 od al punto 2.

Di seguito analizziamo il caso 1 ed il caso 3; del punto 1 vedremo anche l'implementazione tramite il nostro cluster.

### 6.4.1.1. Un Primary + N Secondaries (FailOver Automatico)

Siamo in presenza di una unica macchina che riceve letture e scritture (*Primary*) e di N <= 11 macchine (*Secondaries*) che "possono" essere utilizzate per effettuare delle letture, ma che, per la asincronicità del passaggio delle informazioni e la latenza della rete, potranno essere più o meno allineate al *Primary* e quindi più o meno consistenti.

Con questa configurazione MongoDB garantisce la gestione automatica del fail-over; grazie ad un sistema di elezioni, infatti, le macchine *Secondaries*, o parte di esse, procedono ad una "votazione" al fine di definire democraticamente un nuovo *Primary* nel caso in cui quello in produzione venga a mancare.

Le macchine *Secondaries* possono mantenere i dati o comportarsi solo come elettori fungendo da *arbitri*, in quest'ultimo caso parteciperanno alla votazione permettendo di raggiungere il quorum desiderato concorrendo a raggiungere il numero massimo di *Secondaries* concessi.

Per essere più precisi, le macchine *Secondaries* si dividono in:

- macchine che mantengono le repliche dei dati del *Primary*.

- macchine che partecipano solo alla votazione legata al fail-over.

Per analizzare questa tipologia di configurazione prendiamo le nostre 3 macchine virtuali su cui abbiamo installato le istanze di *MongoDB* e strutturiamo i file di configurazione in modo tale da predisporre una replica tra le tre istanze, o meglio tra il *Primary* e i i due *Secondaries*.

```
Server 1: 192.168.56.220:27017
Server 2: 192.168.56.221:27018
Server 3: 192.168.56.222:27019
```

*Figure 6.4.1 - Istanze MongoDB in replica*

Possiamo rappresentare la struttura che stiamo creando come segue:

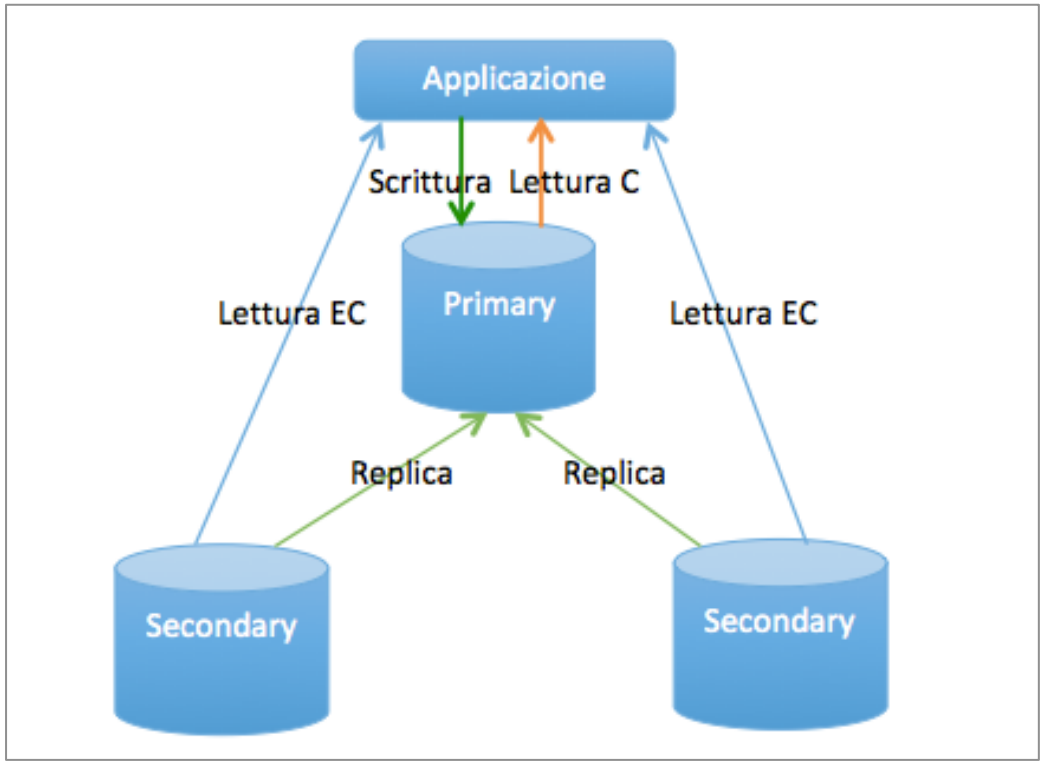

*Figure 5.4.2 – Rappresentazione di un cluster Primary – Secondaries con letture eventualmente consistenti.*

Come si vede l'applicazione parla solo e sempre con la macchina *Primary* per effettuare le scritture e le letture che devono garantire consistenza dei dati, mentre le letture possono avvenire dalle macchine *Secondaries* ma in tal caso non abbiamo la certezza che l'informazione sia sempre consistente (si parla di letture eventualmente consistenti).

### 6.4.1.2. N mongos + M Primary + S Secondaries (Sharding)

Per splittare i dati su più macchine ed avere più di un solo *Primary* atto a mantenere le informazioni, viene usato lo *sharding*.

In MongoDB lo sharding viene garantito tramite l'uso di macchine ad esso dedicate dette *mongos*.

Queste macchine non contengono dati (a differenza delle macchine *mongod*) ma metadati che permettono di accedere in maniera corretta agli N cluster di macchine *mongod* (passando per ognuno di questi cluster sempre e comunque dal *Primary*).

*MongoDB* si accolla la gestione dello sharding una volta che questo è stato configurato opportunamente: per intenderci, una volta scelta la chiave di sharding, le istanze *mongos* accedono alle macchine dei dati (*mongod)* correttamente perché hanno tutte le informazioni necessarie sul dove trovare i dati richiesti e la macchine *Primary* da contattare per scrivere e leggere consistentemente.

Per garantire un effettivo sharding ed un fail-over, servono almeno 2 macchine *mongos* e un numero minimo di 2 cluster ognuna costituito da 3 istanze costruite come al paragrafo 6.4.1.1.

Rappresentiamo graficamente la struttura nel suo complesso, simulando due scritture, una arancio su un Primary e una verde sull'altro Primary passando sempre dalla stessa istanza *mongos*.

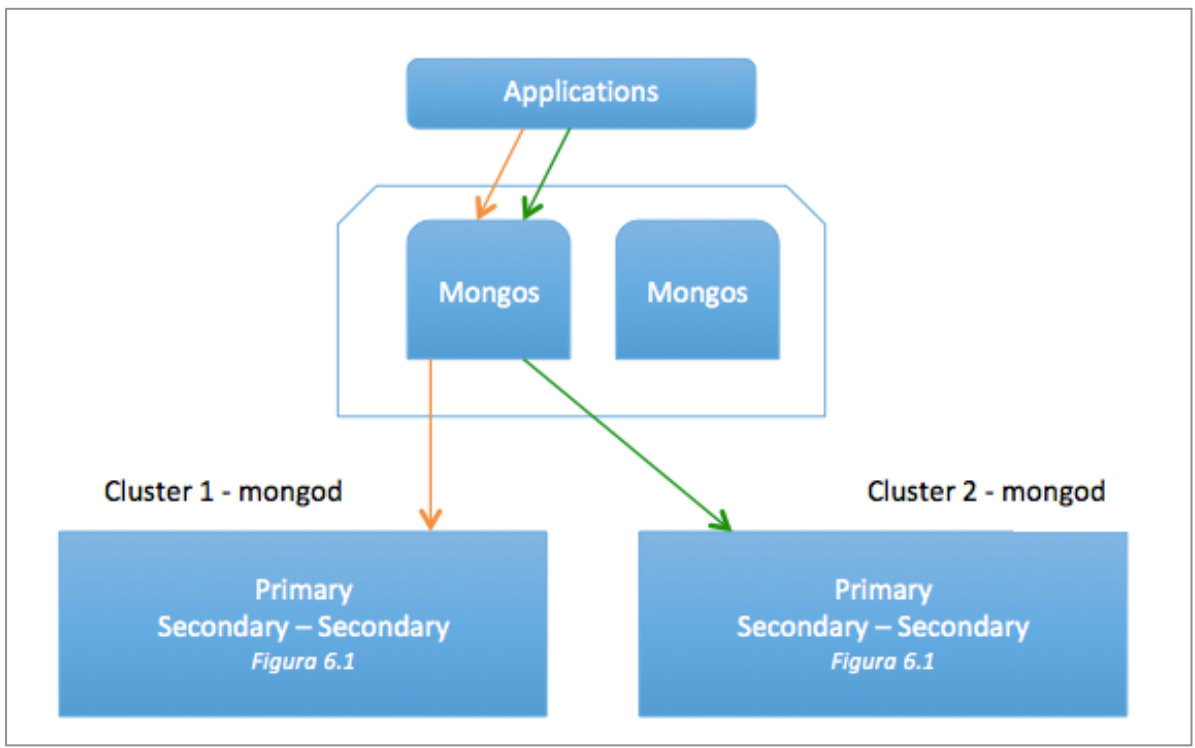

*Figura 6.4.3 – Sharding con failover per MongoDB : l'applicativo effettua una lettura (od una scrittura) di una informazione ad una macchina mongos, che si accolla l'onere di verificare dove essa si trova (o dove andrà scritta), recuperarla (o scriverla) e ritornarla (o comunicare l'avvenuta scrittura) all'applicativo. L'applicativo non saprà dove le informazioni risiedono ma sarà sempre il layer formato dai server mongos a far da filtro per accedere alla corretta macchina che mantiene l'informazione desiderata.*

## 6.4.3. Riak

In Riak i concetti di Primary/Secondaries o Master/Slave non esistono, si parla di *ring* o *anello* o *cluster* e di nodi costituenti.

In Riak un nodo che appartiene ad un cluster se viene interrogato può :

- Fornire direttamente le informazioni richieste.

    *Le informazioni sono presenti sul nodo.*

- Recuperare le informazioni per poi rispondere alla richiesta.

    *Le informazioni sono su un altro nodo e quindi vengono richieste dal nodo contattato al nodo che le mantiene.*

- Verificare che la richiesta sia soddisfatta e fornire le informazioni desiderate.

    *Le informazioni devono essere garantite con un certo grado di affidabilità e quindi vengono contattati più nodi che contengono le medesime e se i dati sono coerenti allora si da una risposta positiva alla richiesta (il concetto di "quorum").*

La prima cosa da notare è che siamo in presenza di un sistema distribuito e decentralizzato dove non ci sono ulteriori macchine se non quelle che appartengono al cluster e che contengono i dati. Sono queste macchine stesse che si occupano di recuperare le informazioni richieste, ovunque esse siano nel cluster.

E' chiaro da questa affermazione che *Riak* fornisce un sistema di sharding automatico; un vantaggio notevole se si pensa che i nodi stessi che mantengono le informazioni ridistribuiscono le medesime sopra di essi sapendo dove e come recuperarle; vantaggio che appare molto più ampio se si pensa alla semplicità della gestione all'aggiunta di un nodo o alla fuoriuscita di uno dal cluster.

Per sapere dove i dati si trovano, ogni nodo usa un *protocollo di gossip*: quest'ultimo permette ad ognuno di essi di mantenersi aggiornato e di sapere, se contattato dall'applicativo, dove recuperare i dati che non ha.

La replica viene gestita allo stesso modo: saranno sempre i nodi, tramite il meccanismo di *consistent hashing*, a mantenere automaticamente il posizionamento dei dati sui nodi del cluster ricollocandoli correttamente se uno o più nodi dovessero abbandonare o venire aggiunti all'anello.

La replica in *Riak* può essere configurata diversamente a livello di *bucket*, ovvero: il numero di nodi su cui i dati vengono replicati può cambiare su ogni *bucket*.

Riak ha dei limiti, tra cui:

- Non è possibile modificare la dimensione di un cluster, ovvero il numero di partizioni che lo formano, a meno di inserire tutti i dati su un nuovo cluster, quindi smantellando un cluster per uno nuovo (il che significa mantenere due strutture fino a quando tutti i dati non sono stati effettivamente trasferiti).

- Il cambio di metodo di archiviazione dei dati, ovvero il *backend*, non può essere fatto se non reinserendo i dati in un nuovo cluster con il backend desiderato.

- Il backend deve essere lo stesso per tutti i nodi del cluster.

Vediamo ora come Riak si adatti al nostro cluster di 3 istanze.

Lo spazio messo a disposizione per le chiavi è di *2^160-bit* di interi e di default sono 64 le partizioni su cui questo spazio è suddiviso.

Avendo noi 3 nodi le partizioni non si distribuiranno equamente per ognuno di questi essendo il modulo di 64/3 = 1. Quindi un nodo avrà una partizione in più.

```
Nodo 1 : 22 partizioni
Nodo 2 : 21 partizioni
Nodo 3 : 21 partizioni
```

*Figure 6.4.4 – Distribuzione delle partizioni tra i nodi*

In Riak, però, si parla di nodi virtuali (*vnodes*), quindi supponendo di avere 3 nodi virtuali per ogni nodo si avrà, ad esempio, per il nodo 1:

```
Nodo 1 : 22 partizioni
      VNode1 : 8 partizioni
      VNode2 : 7 partizioni
      VNode3 : 7 partizioni
```

*Figure 6.4.5 – Distribuzione delle partizione tra nodi virtuali su un singolo nodo*

e così via per gli altri.

Graficamente possiamo meglio vedere questa distribuzione.

Rappresentiamo un cluster di 24 partizioni per comodità di visualizzazione

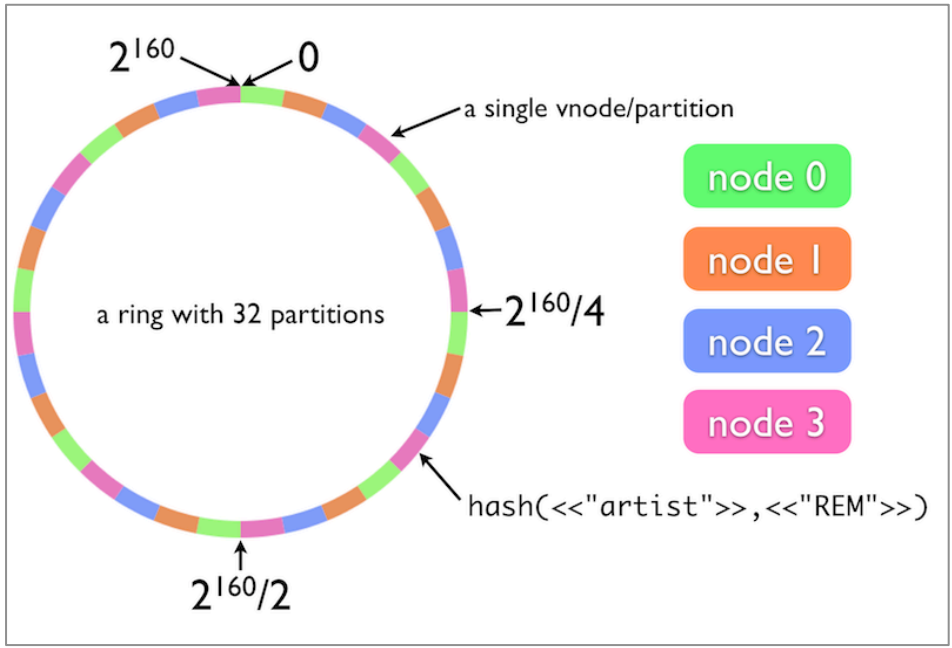

*Figure 6.4.6 – Cluster Riak : distribuzione dell'informazione*

I dati vengono distribuiti lungo le partizioni in maniera equa usando un meccanismo di *consistent hashing* che codifica **bucket + chiave** e trova la partizione che deve contenere il dato o da cui recuperare il dato.

Il nostro cluster in Riak è formato dalle 3 macchine virtuali che formano l'anello:

```
bash> riak-admin status | grep members
ring_members : ['riak@riak1.riak.com','riak@riak2.riak.com',
                'riak@riak3.riak.com']
```

*Figure 6.4.7 - Verifica anello Riak*

## 6.4.4. Cassandra

Anche per Cassandra non esiste il concetto di master, ogni nodo del cluster è uguale agli altri fatta eccezione alcuni fra essi a cui viene dato il ruolo di *seeds.* Questi ultimi devono essere fatti partire prima di tutti gli altri che costituiranno il cluster ed hanno il compito di trovare i nodi mancanti e completare l'anello.

Fatta questa premessa, una volta che l'anello è in piedi ogni nodo è uguale all'altro; quando viene fatta una richiesta ad un nodo è il medesimo che sa dove recuperare le informazioni richieste o dove memorizzare quelle inviate.

Cassandra garantisce scalabilità orizzontale e lineare, il che permette di aggiungere o togliere nodi al cluster di partenza sia per estendere che per migliorare le prestazioni del cluster: questa operazione può essere fatta a caldo, senza restartare il cluster.[c]

---

[c] *Con scalabilità lineare intendiamo che se due nodi possono sostenere 100.000 transazioni al secondo, 4 ne potranno sostenere fino a 200.000 e 8 fino a 400.000.*[12]

I file di configurazione di Cassandra vengono scritti di default con *yalm* che ha la caratteristica di essere molto di più facile lettura che xml.

Nel nostro caso tutti sono dei nodi *seeds*, quindi dopo aver configurato correttamente la replica nei vari file di configurazione possiamo vedere lo stato della stessa con il comando *nodetool* a disposizione tra i comandi bash di Cassandra.

A differenza di MongoDB e Riak, la replica in Cassandra viene configurata esclusivamente nel file di configurazione ed allo start le macchine si "cercano" e trovano senza fare nient'altro.

Di seguito i parametri di configurazione base cluster (ad esempio per il nodo 192.168.56.223) e il recupero dello stato del cluster una volta avviato.

```
cluster_name: 'MyCluster'
num_tokens: 256
seed_provider:
  - class_name: org.apache.cassandra.locator.SimpleSeedProvider
  - parameters:
    - seeds:  "192.168.56.223, 192.168.56.224, 192.168.56.223"
listen_address: 192.168.56.223
rpc_address: 0.0.0.0
endpoint_snitch: RackInferringSnitch
```

*Figure 6.4.8 – Spaccato di un file di configurazione per un singolo nodo per generare un cluster*

```
[root@cassandra1 cassandra]# ./bin/nodetool status
Datacenter: 168
===============
Status=Up/Down
|/ State=Normal/Leaving/Joining/Moving
--  Address         Load       Tokens  Owns   Host ID                               Rack
UN  192.168.56.225  65.4 KB    256     67.0%  97161777-f161-414b-8846-88d1e57144a1  56
UN  192.168.56.223  65.33 KB   256     65.5%  9980b89d-4e05-4623-bfd3-2caa3c177880  56
UN  192.168.56.224  65.31 KB   256     67.4%  b4d30a64-37eb-4a93-b424-f83ca074da30  56
```

*Figure 6.4.9 - Stato del cluster, istanze in replica e sharding dei dati.*

## 6.4.5. Neo4j

In Neo4j torniamo ad una struttura *Master/Slave* senza la possibilità di avere sharding dei dati: il *Master*, come tutti gli *Slave* in replica, conterrà tutti i dati appartenenti all'intero grafo.

Vediamo di seguito i parametri principali per garantire HA nel nostro solito cluster di macchine virtuali.

I file da configurare sono i seguenti:

- conf/neo4j.properties
- conf/neo4j-server.properties

ed i parametri per garantire una replica:

- ha.server_id

  *identificativo univoco per ogni istanza appartenente al cluster.*

- ha.initial_hosts

  *ip e porta per ogni host appartenente al cluster.*

- org.neo4j.server.webserver.address

  *indirizzo ip di risposta del webserver, 0.0.0.0 permette a chiunque l'accesso.*

org.neo4j.server.database.mode

  *abilita/disabilita meccanismo di HA*

Una volta configurati questi parametri, avviando le istanze, null'altro verrà chiesto per rendere disponibile la replica.

Vediamo di seguito la configurazione di un macchina del nostro cluster.

```
ha.server_id = 1
ha.initial_hosts = 192.168.56.223:5001,
                   192.168.56.224:5001,
                   192.168.56.225:5001
```

*Figure 6.4.10 - neo4j.properties : spaccato per la creazione di un cluster*

```
org.neo4j.server.webserver.address = 0.0.0.0
org.neo4j.server.database.mode = HA
```

*Figure 6.4.11 - neo4j-server.properties : spaccato per la creazione di un cluster*

Si potrà vedere, poi, lo stato della replica su ogni macchina tramite browser; in *Figura 6.4.10* è mostrata parte della pagina delle informazioni a disposizione nella sezione HA.

*Figure 6.4.12 - HA dal master : il parametro InstancesInCluster mantiene le informazioni per tutte le istanze appartenenti al cluster.*

## 6.5. Connessione, accesso ai dati, caricamento massivo.

L'analisi che segue è stata fatta con repliche già configurate per tutti i NoSQL, quindi i tempi di preparazione dei dati e propagazione delle informazioni per la replica sono da considerarsi già inseriti nei valori recuperati.

In seguito vedremo:

- Come interagire da console con ogni database.

- Grazie alla creazione (se serve) di una struttura dati ad hoc (*Appendice B*) ed al suo popolamento, come interfacciarsi con i dati inseriti al fine di ottenere raggruppamenti di informazioni, somme, relazioni o quant'altro.

- Dei caricamenti massivi di dati usando dei driver creati ad hoc per Python al fine di interfacciarsi con ogni NoSQL esaminato. I dati sintetici sono generati da un apposito script (*Appendice C*).

## 6.5.1. MongoDB

### 6.5.1.1. Accesso alla console e generazione dei dati

MongoDB è dotato di una console accessibile via bash tramite il comando _mongo_ che permette di interrogare i database storati ed i loro contenuti grazie a funzioni javascript create ad hoc.

```
bash# mongo
MongoDB shell version: 2.4.3
set0:PRIMARY>
```

*Figure 6.5.1 – Connessione alla console*

Vediamo anzitutto come interrogare una macchina _mongod_ nel nostro cluster.

Solo una macchina alla volta può essere Primary indipendentemente dalla configurazione prescelta; tutte le altre sono Secondaries. Questo evidenzia il fatto che una macchina avrà possibilità di read/write consistenti (il Primary) mentre le altre permetteranno solo read eventualmente consistenti (le Secondaries).

Nel nostro cluster, senza macchine _mongos,_ per effettuare un inserimento di una informazione da console, dobbiamo prima accertarci di quale sia la macchina Primary; il comando che ci permette di farlo è il seguente:

```
set0:SECONDARY> db.isMaster().primary
192.168.56.223:27017
```

*Figure 6.5.2 - Recupero informazioni relative al Primary.*

Che ci ritorna IP e PORTA del Primary; a questo punto ci basta salire sulla macchina identificata ed effettuare le scritture desiderate.

```
mypc> ssh 192.168.56.223
bash# mongo
MongoDB shell version: 2.4.3
set0:PRIMARY>
```

*Figure 6.5.3 - Collegamento al Primary.*

### 6.5.3.2. Accesso ai dati

Con riferimento all'esempio riportato in Appendice B, genereremo un nuovo *database: automobili* e raccoglieremo le informazioni in due *collections : autovetture* e *produttori*, dalle quali poi andremo a recuperare i dati desiderati.

La creazione di una *collection* avviene automaticamente introducendo dati in essa.

Il codice nella *figura 6.5.4* rappresenta la situazione dopo gli inserimenti, che sono proposti in Appendice; qui visualizziamo i comandi per vedere database e collections per esso.

```
set0:PRIMARY> use automobili
switched to db automobili
set0:PRIMARY> show collections;
autovetture
produttori
system.indexes
```

*Figure 6.5.4 – Visualizzazione delle collections appartenenti al database automobili.*

Le caratteristiche principali di MongoDB nell'accedere ai dati sono:

- ogni singola operazione di scrittura non può interessare più di una *collection*.
- tutte le operazioni di scrittura di MongoDB sono *atomiche* a livello di singolo documento.

Passiamo ora ai comandi principali per interagire con le collections appena create.

*Cancellazione del contenuto di una collection*

In MongoDB la rimozione completa delle informazioni dentro una collections è possibili grazie a due comandi :

- *drop*: con questa cancellazione vengono dismessi contenuto e collection.
- *remove*: con questa cancellazione il contenuto della collection viene rimosso ma la collection viene mantenuta.

*Filtro su un valore di una chiave*

Il comando *find* ci permette di visualizzare i documenti inseriti nella collection; per vedere invece una parte di queste informazioni è necessario passare al comando la chiave ed il valore associato che si desidera recuperare dai documenti, i documenti che non vedranno verificata la condizione saranno scartati.

Di seguito un esempio di filtro su un dato campo.

```
set0:PRIMARY> db.autovetture.find({'modello':'Punto'})
{ "_id" : 1, "marca" : "Fiat", "modello" : "Punto", "tipologia" :
"Utilitaria", "alimentazione" : "Benzina", "cilindrata" : "1400", "prezzo" :
10 }
{ "_id" : 2, "marca" : "Fiat", "modello" : "Punto", "tipologia" :
"Utilitaria", "alimentazione" : "GPL", "cilindrata" : "1200", "prezzo" : 12 }
```

*Figure 6.5.5 – Recupero dei documenti filtrati sul campo/valore scelto*

## Join di informazioni

La join in MongoDB non è supportata, serve quindi normalizzare i dati duplicando le informazioni nelle collections che si vorrebbero unire.

Nel nostro caso il campo marca ci permette di sapere, data una autovettura, dove ha sede il produttore dell'autovettura stessa tramite due separate *queries*.

```
set0:PRIMARY> db.autovetture.find({'modello':'Punto','alimentazione':'GPL'})
{ "_id" : 2, "marca" : "Fiat", "modello" : "Punto", "tipologia" :
"Utilitaria", "alimentazione" : "GPL", "cilindrata" : "1200", "prezzo" : 12 }
set0:PRIMARY> db.produttori.find({'marca':'Fiat'},{'marca':1,'citta':1})
{ "_id" : 1, "marca" : "Fiat", "citta" : "Torino" }
```

*Figure 6.5.6 – Join tramite normalizzazione del campo marca.*

## Raggruppamento di informazioni e conteggio

Per raggruppare le informazioni dobbiamo servirci di un sistema di *Map-Reduce* che ci permette di mappare le informazioni che ci interessano nei json rappresentanti i documenti (nell'esempio sottostante prenderemo tutti i documenti con campo "alimentazione") riducendoli a un elenco contenente le occorrenze per ogni insieme trovato.

```
set0:PRIMARY> db.autovetture.group( {key:{'alimentazione':1},
                                     reduce:function(curr,result)
                                        {result.total +=1},
                                     initial: { total:0}} );
[
     {      "alimentazione" : "Benzina",
            "total" : 3 },
     {      "alimentazione" : "GPL",
            "total" : 1 },
     {      "alimentazione" : "Diesel",
            "total" : 4 }
]
```

*Figure 6.5.7 - Raggruppamento per alimentazione, MapReduce.*

### 6.5.1.3. Inserimento massivo

Per verificare i tempi di inserimento di informazioni tramite un caricamento massivo dei dati usiamo pymongo, un driver Python per MongoDB.

I tempi di caricamento saranno osservati con uno, dieci o venti thread contemporanei.

Durante il caricamento la CPU è spesso al 100%, la RAM va al 75% mentre l'IO varia con picchi che raggiungono il 95% di carico.

Di seguito la tabella con tutti gli esperimenti riportati.

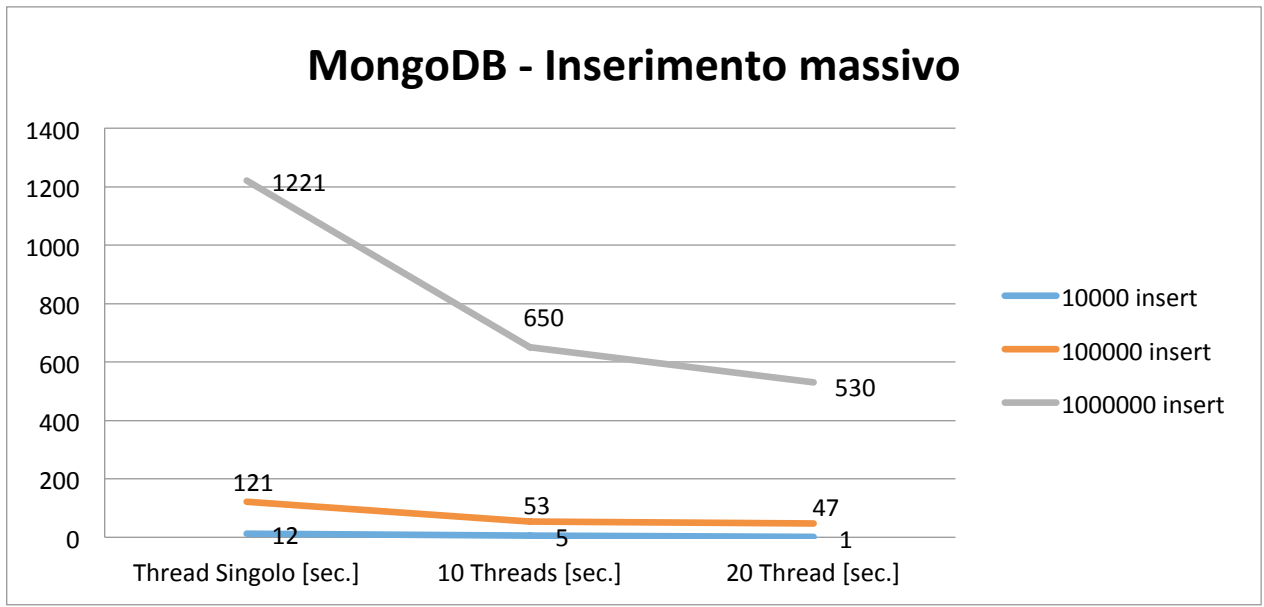

*Figure 6.5.8 - Inserimento dati al cambio del numero di threads - MongoDB*

Si può notare come l'aumento di threads incida notevolmente sulla velocità di inserimento delle informazioni, purtroppo però MongoDB, avendo solo n Primary su cui effettuare gli inserimenti, non può scalare orizzontalmente (ricordiamo che stiamo analizzando il caso di cluster con 3 nodi senza sharding).

## 6.5.2. Riak

### 6.5.2.1. Accesso alla console e generazione dei dati

Per accedere ai dati Riak usa una console *Erlang*, quindi le query da console vengono fatte tramite questo linguaggio.

Riak, nel nostro cluster, gira in background, quindi per accedere alla console useremo il comando *riak-attach*.

```
bash> riak-attach
Remote Shell: Use "Ctrl-C a" to quit. q() or init:stop() will terminate the
riak node.
Erlang R15B01 (erts-5.9.1) [source] [64-bit] [async-threads:0] [kernel-
poll:false]

Eshell V5.9.1  (abort with ^G)
(riak@riak1.riak.com)1>
```

*Figure 6.5.9 – Accesso alla console di Riak.*

E' possibile interfacciarsi con Riak anche tramite chiamate REST, sia per caricare che consultare i dati.

E' doveroso tenere in considerazione che, per caricare un nuovo dato, Riak necessita di ricevere il suo *Content-Type*, mentre per un update deve essere letto ed aggiornato anche il *Vector-clock* dell'informazione stessa.

### 6.5.2.2. Accesso ai dati

In Riak non c'è la possibilità di creare più database sulle istanze di un cluster, quindi con riferimento all'esempio in Appendice B, nelle nostre istanze esisteranno due *buckets* che corrisponderanno ai nostri insiemi di *autovetture* e *produttori*.

I *buckets* non sono niente altro che dei namespace, ovvero un metodo per avere una distinzione logica delle informazioni.

Quello che Riak ci permette di avere in più a livello di *bucket* è definire dei fattori di replica diversi per ognuno di questi. Quindi potremmo pensare di replicare 3 volte le informazioni per il bucket contenenti le autovetture mentre 2 volte solo quello che raccoglie i produttori.

Questo inciderà sull'avere scritture e/o letture più o meno veloci a causa dei tempi di latenza della rete per la propagazione corretta delle scritture o per il recupero corretto dell'informazione.

Vediamo degli esempi di accesso alle informazioni tramite chiamate API HTTP messe a disposizione da Riak; ricordiamoci che siamo in presenza di un Key-Value e che quindi le operazioni di base sono inserimenti (Create), letture (Read), modifiche (Update) e cancellazioni (Delete), note anche come CRUD.

*Cancellazione completa del contenuto di un bucket*

Questa operazioni non è possibile in Riak a meno di cancellare uno alla volta i dati presenti nel bucket stesso. In alternativa, dopo aver stoppato le istanze, si può procedere alla cancellazione della cartella contenente i dati su tutte le macchine del cluster, riavviando poi le istanze.

*Filtro su un valore di una chiave*

Per effettuare ricerche di un dato valore sui contenuti, essendo Riak un key-value, ci si affida ad una indicizzazione basata su Lucene che permette di eseguire le ricerche con chiamate Solr-like.

Per riempire questo indice Riak usa un *hook* al commit dell'inserimento dell'informazione.

Un esempio di chiamata *Solr-like* per recuperare tutti i modelli "Punto" tra tutti i documenti può essere vista in *figura 6.5.10*, dove vengono mostrati anche i risultati.

```
bash> curl \
"http://192.168.56.223:8098/solr/autovetture/select?start=0&rows=10& /
q='modello:Punto&fl=marca,modello,alimentazione"
<xml>
<doc id=1>
 <marca>Fiat</marca>
 <modello>Punto</modello>
 <alimentazione>Benzina</alimentazione>
</doc>
<doc id=2>
 <marca>Fiat</marca>
 <modello>Punto</modello>
 <alimentazione>Benzina</alimentazione>
</doc>
```

*Figure 6.5.10 - Recupero di un insieme di informazioni tramite indice ad-hoc*

### Raggruppamento di informazioni

In Riak per aggregare le informazioni si utilizza il paradigma di Map-Reduce.

```
curl -H 'Content-Type: application/json' http://192.168.56.223:8098/mapred \
    -d '{
        "inputs":"autovetture",
        "query":[{
            "map":{
                "language":"javascript",
                "source":"function(riakObject) {
                        var m = riakObject.values[0];
                        return m ;
                    }"
            }
        }]
    }'
```

*Figure 6.5.11 – Esempio di chiamata HTTP per Map-Reduce*

## 6.5.2.3. Bulk Insert

Per effettuare un test di inserimento massivo con Riak, usiamo la libreria python-riak-client.

La libreria permette di creare un pool di connessioni e, per ogni chiamata verso Riak, una tra queste connessioni viene usata, random. Questo permette di avere scritture già distribuite sui diversi nodi.

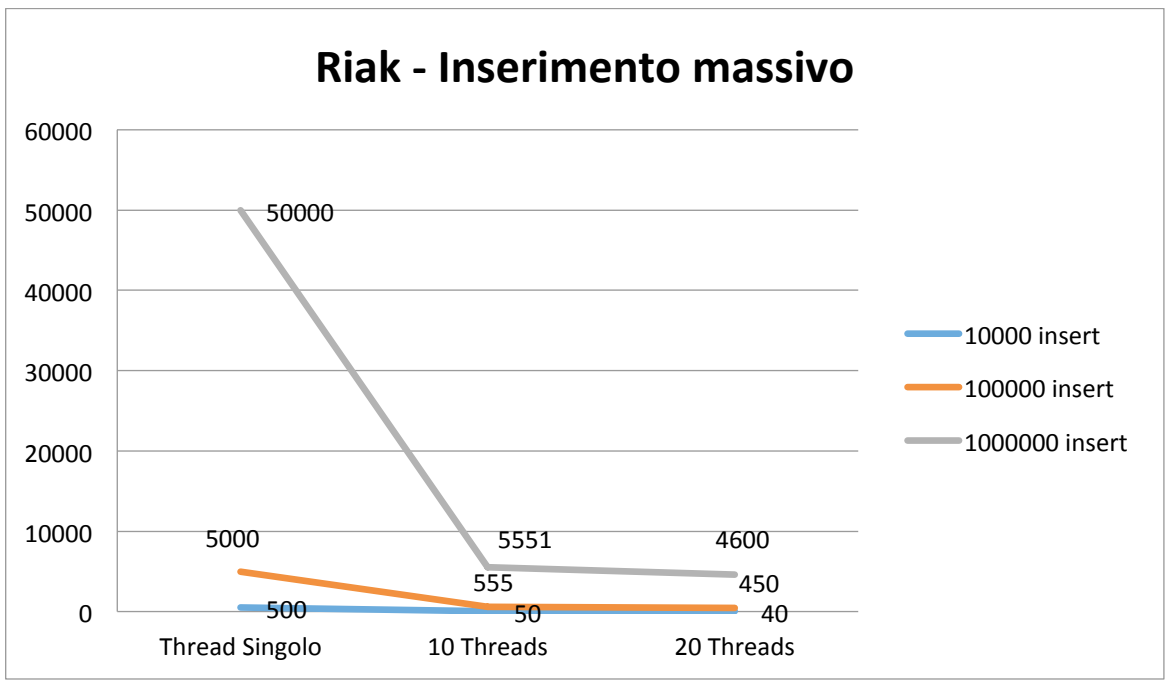

*Figure 6.5.12 – Inserimento dati con differente numero di threads. Riak*

Si può notare come l'inserimento sia lineare all'aumentare dei dati mantenendo costante il numero di thread. Invece si vede un netto miglioramento dal thread singolo ai 10 thread, miglioramento che diventa meno evidente passando da 10 a 20 threads.

## 6.5.3. Cassandra

### 6.5.3.1. Accesso alla console e generazione dei dati

Alla console di Cassandra 2.0 si accede tramite il comando *cql*, acronimo di *Cassandra Query Language*, linguaggio di interrogazione del DBMS.

Supposto che i binari di Cassandra si trovino nella directory */usr/local/cassandra*, vediamo come accedere alla console.

```
bash> cd /usr/local/cassandra
bash> ./bin/cqlsh
Connected to My Cluster at localhost:9160.
[cqlsh 4.1.0 | Cassandra 2.0.2 | CQL spec 3.1.1 | Thrift protocol 19.38.0]
Use HELP for help.
cqlsh>
```

*Figure 6.5.13 – Accesso alla console di Cassandra*

Abbiamo accennato al fatto che Cassandra è un NoSQL per il quale serve una definizione di uno schema, quindi, dopo esserci collegati alla console, introduciamo qui la creazione di un *keyspace* (un database in ambito Cassandra) su cui faremo i nostri test di interrogazione e bulk insert nei prossimi paragrafi.

Il nostro keyspace, inerente all'esempio che si trova in Appendice B, viene creato definendo una *motodologia di replica* ed un *fattore di replica*, ovvero il tipo di replica che si desidera con il numero di repliche che si desidera avere per i dati.

```
cqlsh> CREATE KEYSPACE automobili WITH replication = {
                   'class': 'SimpleStrategy', 'replication_factor': 3}
cqlsh> USE automobili;
```

*Figure 6.5.14 - Creazione ed uso di un keyspace*

Oltre al database anche le tabelle devono essere precedentemente generate; nel nostro caso avremo il keyspace *automobili* avrà le due tabelle *autovetture* e *produttori*.

Creiamo ora le due tabelle, dette anche *Column-Family*.

```
cqlsh:automobili> CREATE TABLE autovetture ( marca varchar,
                                             modello varchar primary key,
                                             tipologia varchar,
                                             alimentazione varchar,
                                             cilindrata varchar,
                                             prezzo int);

cqlsh:automobili> CREATE TABLE produttori ( marca varchar,
                                            citta varchar,
                                            nazione varchar,
                                            email varchar);
```

*Figure 6.5.15 - Column Families per keyspace automobili*

### 6.5.3.2. Accesso ai dati

*Cancellazione completa del contenuto di un bucket*

La cancellazione del contenuto di una tabella è possibile in Cassandra, tramite il comando DELETE.

*Filtro su un valore di una chiave*

Supponiamo di voler recuperare le informazioni di tutte le utilitarie che abbiamo tra le automobili.

Per fare si che una query del genere dia risultati, bisogna creare un indice apposito.

In principio possiamo solo eseguire query sulla chiave primaria, ovvero sulla chiave *modello* (come si può vedere guardando la creazione della tabella *autovetture* in *figura 6.5.15*.

Ecco quindi come creare l'indice sul campo tipologia.

```
cqlsh:automobili> CREATE INDEX tipologia ON autovettue (tipologia) ;
```

*Figure 6.5.16 - Creazione di un ulteriore indice*

Ora che abbiamo l'indice possiamo interrogare la base di dati e vedere tornare i risultati aspettati (*figura 6.5.17*).

```
cqlsh> SELECT * FROM automobili.infobase WHERE tipologia='Utilitaria';

 modello | marca       | motore         | tipologia

---------+-------------+----------------+------------

   Punto |       Fiat  |        Benzina | Utilitaria
   Punto |       Fiat  |            GPL | Utilitaria
      A1 |       Audi  |        Benzina | Utilitaria
    Clio |     Renault |        Benzina | Utilitaria

```

*Figure 6.5.17 – Elenco delle utilitarie tra le autovetture*

### *Raggruppamento di informazioni*

Fare un GROUP BY in gergo MySQL non è possibile con Cassandra, dobbiamo quindi percorrere una strada più onerosa.

Creiamo un altro indice per la marca e per ogni marca vediamo il numero di utilitarie che hanno.

```
cqlsh:automobili> CREATE INDEX marca ON autovetture (marca);

cqlsh:automobili> SELECT count(*) FROM automobili.autovetture
                            WHERE tipologia = 'Utilitaria'
                            AND marca = 'Fiat';
Bad Request: Cannot execute this query as it might involve data filtering and
thus may have unpredictable performance. If you want to execute this query
despite the performance unpredictability, use ALLOW FILTERING
cqlsh:automobili> SELECT count(*) FROM automobili.autovetture
                            WHERE tipologia = 'Utilitaria'
                            AND marca = 'Fiat' ALLOW FILTERING;

 count
-------
     3
(1 rows)
```

*Figure 6.5.18 – Indice sul campo marca e raggruppamento dei dati*

### 6.5.3.4. Bulk Insert

Andiamo ora ad effettuare delle prove di carico tenuto presente che il fattore di replica è pari al numero di nodi massimo che stiamo usando, 3.

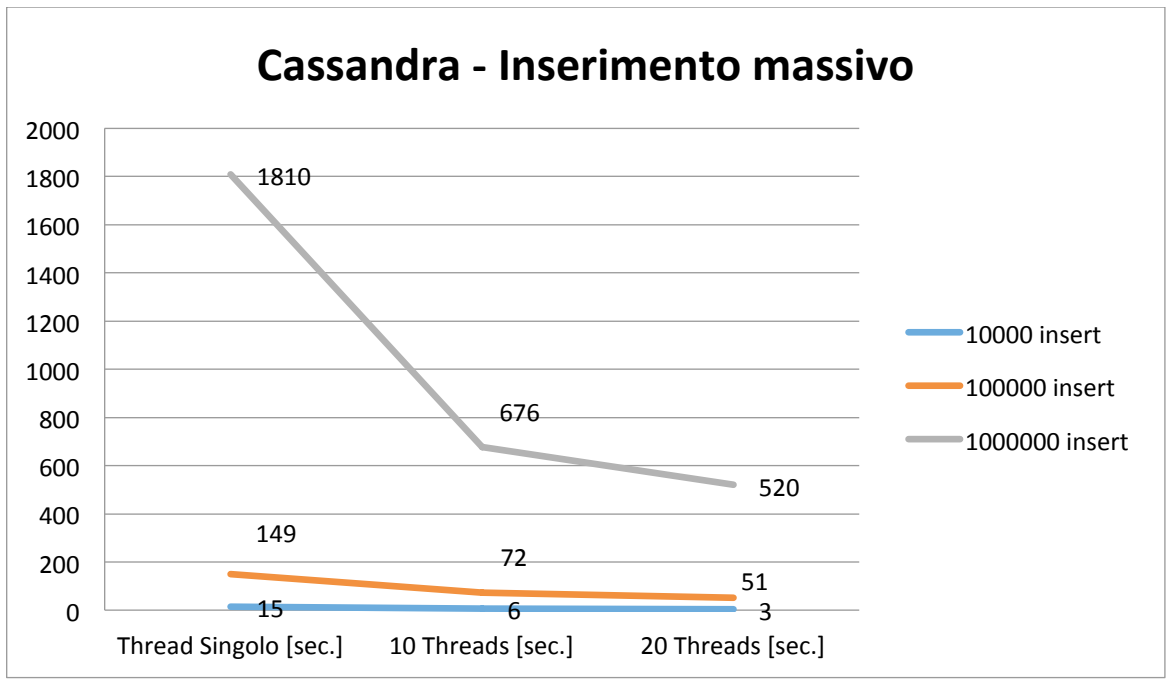

*Figure 6.5.19 - Inserimento dati al cambio del numero di threads – Cassandra*

Durante il caricamento a singolo thread si può vedere sulla macchina un carico si CPU pari al 99.9%, mentre per quanto riguarda la RAM siamo attestati al 70% e IO molto basso, 2%, con punte a 30% ma con macchine virtuali sopra una macchina fisica difficile fare una valutazione se picchi dovuti a questo inserimento o della macchina fisica stessa per altri processi in esecuzione.

## 6.5.4. Neo4j

### 6.5.4.1. Accesso alla console e generazione dei dati

Neo4j usa *Cypher* (*cql : Cypher Query Language*) come linguaggio per inserire o cercare informazioni tramite console; quindi, supposto che Neo4j sia stato installato in */usr/local/neo4j*, avremo:

```
bash> cd /usr/local/neo4j
bash> ./bin/neo4j-shell
Welcome to the Neo4j Shell! Enter 'help' for a list of commands
NOTE: Remote Neo4j graph database service 'shell' at port 1337
neo4j-sh (0)$
```

*Figure 6.5.20 – Accesso alla console di Neo4j.*

### 6.5.4.2. Accesso ai dati

Vediamo ora come inserire i nostri dati di test relativi al solito esempio di *Appendice B*.

In Neo4j la gestione delle informazioni è piatta, quindi non si possono creare più database o keyspace, tutti gli elementi sono nodi o relazioni corredati da proprietà.

Nel caso che stiamo analizzando riportiamo un grafico ulteriore a sottolineare come le informazioni siano interconnesse evidenziando i dati intesi come nodi e i dati intesi come relazioni fra di essi.

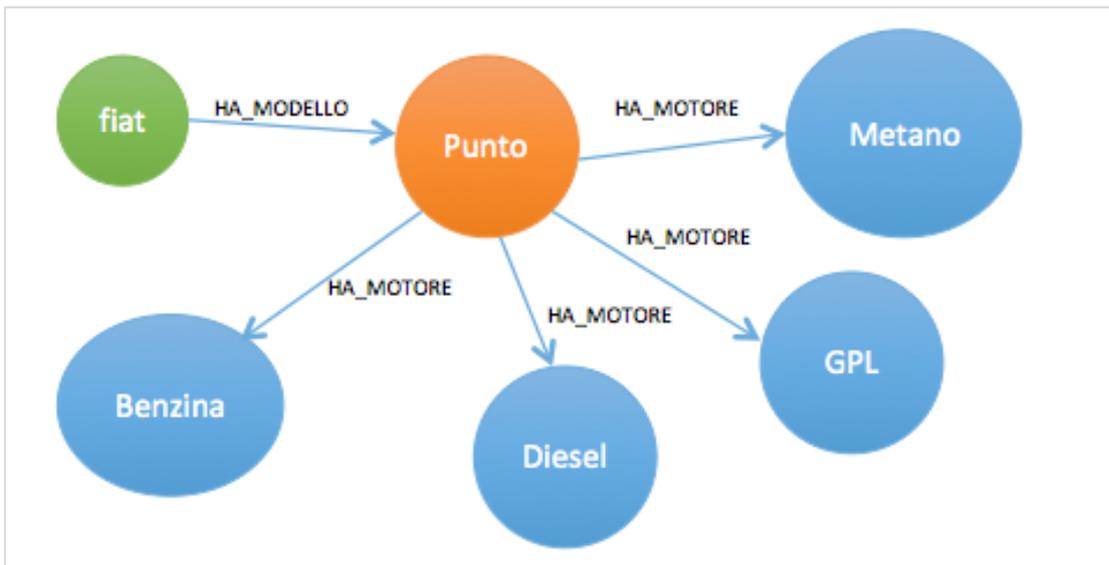

*Figure 6.5.21 - Grafo interconnessioni nodi e relazioni.*

Nella foto i nodi sono evidenziati con dei tondi e, a seconda della categoria, con colori differenti. Le relazioni invece sono frecce che, in questo caso, definiscono una "appartenenza"; in questo esempio che la marca "Fiat" ha un modello che si chiama "Punto" che può avere un motore a "Metano" o a "GPL" o "Diesel" o a "Benzina".

Fatto l'inserimento, che possiamo trovare sempre in *Appendice*, per vedere l'elenco di informazioni, suddividendo *autovetture* e *produttori,* eseguiamo le query riportate nelle prossime due figure.

```
neo4j-sh (?)$ start n = node(*) WHERE has(n.modello) return
n.marca,n.modello,n.alimentazione,n.prezzo;
+--------------------------------------------------+
| n.marca   | n.modello | n.alimentazione | n.prezzo |
+--------------------------------------------------+
| "Fiat"    | "Punto"   | "Benzina"       | 10000    |
| "Fiat"    | "Punto"   | "GPL"           | 12000    |
| "Audi"    | "A1"      | "Benzina"       | 18000    |
| "Audi"    | "A4"      | "Diesel"        | 30000    |
| "Bmw"     | "Serie1"  | "Diesel"        | 24000    |
| "Bmw"     | "Serie3"  | "Diesel"        | 36000    |
| "Renault" | "Clio"    | "Benzina"       | 11000    |
| "Renault" | "Scenic"  | "Diesel"        | 18000    |
+--------------------------------------------------+
```

*Figure 6.5.22 - Elenco delle autovetture inserite visualizzate marca, modello, alimentazione e prezzo*

```
neo4j-sh (?)$ start n = node(*) WHERE has(n.email) return n.marca, n.citta,
n.email;
+------------------------------------------------------------+
| n.marca   | n.citta               | n.email             |
+------------------------------------------------------------+
| "Fiat"    | "Torino"              | "info@fiat.it"      |
| "Audi"    | "Ingolstadt"          | "info@audi.de"      |
| "Bmw"     | "Monaco di Baviera"   | "info@bmw.de"       |
| "Renault" | "Boulogne-Billancourt"| "info@renault.fr"   |
+------------------------------------------------------------+
```

*Figure 6.5.23 - Elenco dei produttori visualizzando marca, città e email*

### Cancellazione completa del contenuto del grafo

Per cancellare interamente il grafo in Neo4j si agisce direttamente sulla directory su file system che lo contiene, avendo la accortezza di ripetere l'operazione per tutte le macchine appartenenti al cluster e soprattutto a istanze stoppate.

### Filtro su un valore di una chiave

Filtrare su un dato valore è molto simile alla sintassi SQL dove si usa la WHERE per ottenere un certo tipo di elementi.

Vediamo di seguito un filtro sul costo autovetture tale che sia minore di una certa soglia.

```
neo4j-sh (0)$ start n = node(*) WHERE has(n.modello)  and(n.prezzo < 20000 )
return n.marca,n.modello,n.alimentazione,n.prezzo;
+------------------------------------------------+
| n.marca   | n.modello | n.alimentazione | n.prezzo |
+------------------------------------------------+
| "Fiat"    | "Punto"   | "Benzina"       | 10000    |
| "Fiat"    | "Punto"   | "GPL"           | 12000    |
| "Audi"    | "A1"      | "Benzina"       | 18000    |
| "Renault" | "Clio"    | "Benzina"       | 11000    |
| "Renault" | "Scenic"  | "Diesel"        | 18000    |
+------------------------------------------------+
```

*Figure 6.5.24 - Autovetture che costano meno di 20000 euro*

### Join di informazioni

Neo4j permette di relazionare le informazioni inserendo appropriatamente le relazioni tra di esse, Questo permette di esplicitare la relazione, non dovere normalizzare le informazioni e quindi non avere delle duplicazioni.

Vediamo come inserire una relazione, ad esempio che "Fiat" fa le "Punto".

```
neo4j-sh (0)$ start n = node(1002) match (n)--(x) return
n.marca,n.citta,x.modello,x.alimentazione;
+------------------------------------------------+
| n.marca | n.citta  | x.modello | x.alimentazione |
+------------------------------------------------+
| "Fiat"  | "Torino" | "Punto"   | "Benzina"       |
| "Fiat"  | "Torino" | "Punto"   | "GPL"           |
+------------------------------------------------+
```

*Figure 6.5.25 - Relazione tra produttore e autovetture prodotte.*

Questo ultimo esempio mostra come la normalizzazione in Neo4j sia inutile e quindi, nel nostro esempio, il campo "marca" dai modelli di autovetture, una volta introdotte le relazioni non servirà più per ogni tupla, ma verrà collegata ai modelli della marca.

Così come il campo marca, anche il campo modello non si ripeterà per modelli uguali con diverse motorizzazioni, ma il modello sarà relazionato a più motorizzazioni, generando la situazione finale rappresentata nella *figura 6.5.21* mostrata all'inizio del paragrafo.

### 6.5.4.3. Bulk Insert

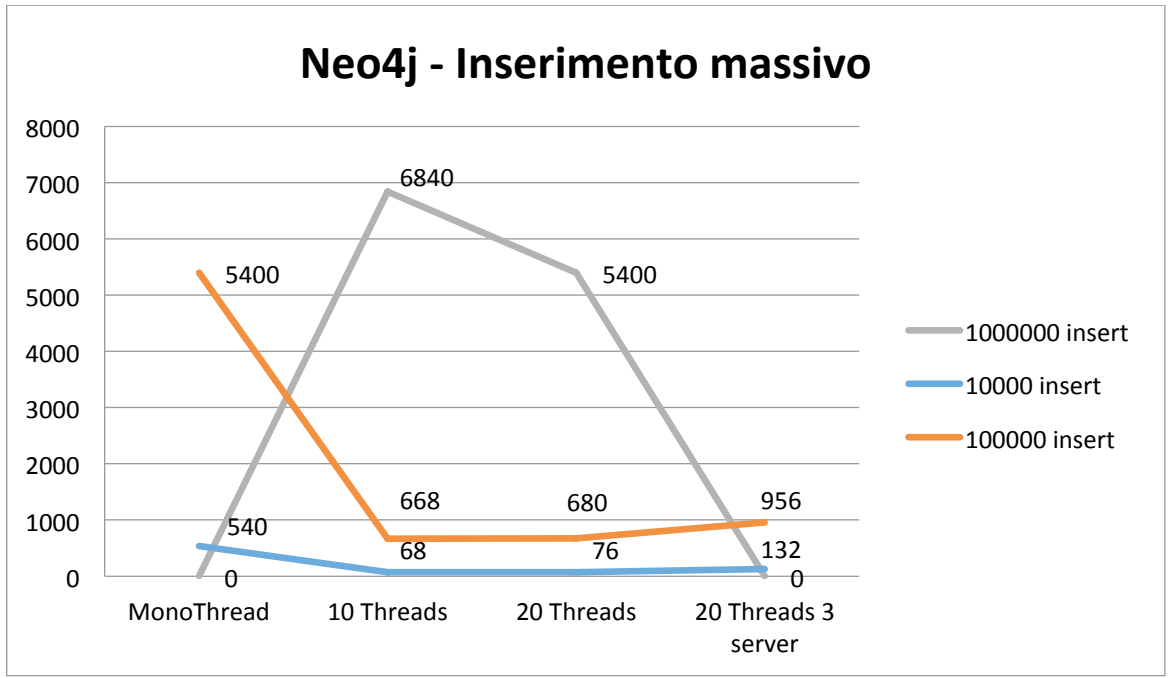

*Figure 6.5.26 – Inserimento dati tramite numero di thread variabile. Neo4j*

Analizzando i dati si osserva che il caricamento tramite mono thread passando verso una sola macchina quindi, è lineare e se per inserire 10000 nodi ci vogliono 9 minuti, per inserirne 100000 ce ne vogliono 90 di minuti: possiamo quindi stimare che per inserirne un milione ci servono 900 minuti ovvero 15 ore.[d]

Se usiamo 10 threads, sempre passando da un solo server per effettuare l'inserimento, servono 68 secondi per fare 10000 insert e 670 circa per farne 100000.

Circa gli stessi dati anche se poco peggiori per 20 threads.

Proseguendo vediamo una quarta voce nel grafico, ovvero la distribuzione dei 20 thread sulle 3 macchine in maniera bilanciata:

- 7 sul primo server

- 7 sul secondo server

- 6 sul terzo server

Questa misurazione è importante in quanto Neo4j, anche se ha una struttura Master/Slave, permette che le scritture avvengano su tutte le macchine e non su una soltanto (a differenza ad esempio di MongoDB).

Salta all'occhio che non si ha un vantaggio per la velocità di inserimento ma chiaramente abbiamo la possibilità di scalare e distribuire il carico in maniera più equilibrata.

In questo caso però i dati di replica che viaggiano sulla rete sono più bassi del 33% circa (avendo un caricamento bilanciato sulle macchine e quindi dovendo ogni macchina ricevere le parti mancanti); le macchine si riallineranno un po' più velocemente del caso del caricamento massivo su una sola macchina: ciò significa avere una consistenza dei dati più rapida ma da contraltare il fatto che i dati mentre si inseriscono non saranno su nessun server consistenti (avendo distribuito il carico e quindi non essendoci un singolo server di scrittura).

---

[d] *Quest'ultimo valore non è riportato graficamente per non "schiacciare" le altre informazioni che ci interessano maggiormente.*

I dati sono copiati interamente sulle altre macchine quindi non c'è uno sharding ma ogni macchina che appartiene al cluster mantiene tutti i dati.

## 6.6. Backup dei dati

Come effettuare il backup e valutazione di tempi per effettuarlo.

## 6.7. Tabella Comparativa Complessiva

|  | MongoDB | Riak | Cassandra | Neo4J |
|---|---|---|---|---|
| Scalabilità Orizzontale | SI | SI | SI | NO? |
| Replica Automatica | NO | SI | SI | NO |
| Presenza nodo Master | SI | NO | NO | SI |
| Scrittura solo su Master | SI | ---- | ---- | NO |
| Schema | NO | NO | SI | NO |
| Livello di Replica | Istanza | Bucket | Keyspace | Istanza |
| Minimo nodi consigliati | 8* | 5 | 3 | 3 |
| Numero file consigliato | ---- | 4096 | ---- | 40.000 |
| Configurazione replica | CF + CL | CF + CL | CF | CF |
| File di configurazione | XML | XML | YAML | XML |
| Struttura gerarchica** | Databases -> Collections -> Data | Buckets -> Data | Keyspaces -> Column Families -> Data | Grafo -> Nodi |
| Interfaccia grafica web | SI | NO | SI | SI |
| Console | JavaScript | Erlang | Cql | Cypher |

** : livello astrazione dei dati; Es. MongoDB : Databases, Collections per ogni database

* : con riferimento alla versione con Sharding :2 mongos + 3x2 mongod

"CF" : File di Configurazione – "CL" : Command Line

# 7. Conclusioni

Dai dati che abbiamo sopra esposto, possiamo affermare che i database NoSQL non vengono scelti in base al gusto personale ma in base alle caratteristiche che devono avere.

Ad esempio, i sistemi con *sharding* non automatico, vedi ad esempio *MongoDB*, fanno si che sia l'utente o comunque l'applicativo a ridistribuire il carico delle informazioni tra i servers che costituiscono l'ambiente. Questo è un compito tedioso e difficile che può portare ad analisi molto approfondite soprattutto se i dati cambiano continuamente nel database.

Per questo motivo sono da preferirsi, nel caso si voglia effettuare uno *sharding*, i sistemi che lo integrano automatico e che solitamente usano il sistema di *consistent hashing*: tra i nostri NoSQL con questa caratteristica troviamo *Riak* e *Cassandra*, mentre Neo4j non è neanche da tenersi in considerazione non supportando la distribuzione tra i nodi.

D'altra parte *MongoDB* è molto più veloce dei diretti concorrenti e permette di effettuare molte tipologie di queries sui dati grazie all'implementazione di parecchi indici di ricerca e alla struttura di *journaling*.

Un'altra caratteristica che salta all'occhio è che *MongoDB*, come *Cassandra*, sono NoSQL multi databases, quindi nel caso di sharding per più database sono da preferirsi a *Riak* e *Neo4j*.

Altro fattore molto importante è verificare se il DBMS NoSQL sia *distribuito e decentralizzato*. *MongoDB* è distribuito ma, avendo un legame 1:1 tra inserimento informazione e *Primary*, non può certo dirsi decentralizzato, caratteristica che hanno in comune gli altri NoSQL esaminati.

*Cassandra* è consigliato inoltre per sistemi che scalano orizzontalmente, ovvero per sistemi per i quali si chiede l'aggiunta di nodi continuamente, e che necessitano di ampliare lo spazio di memorizzazione dell'informazione nel tempo; inoltre la sua suddivizione per tipologia delle informazioni permette di avere un migliore accesso a sistemi che necessitano recuperi di informazione omogenea.

Schematicamente.

**MongoDB**

PRO

- Usa molto disco ma è molto performante.

- Da usare per caricamenti continui e veloci e se sono da gestire più di un database.

- Lo sharding permette di splittare i dati ma più copie degli stessi dati appartengono sempre e comunque a ad un sottoinsieme *Primary/Secondaries* esclusivo.

- Failover garantito.

- Ha macchine specifiche che si occupano di fare lo sharding.

CONTRO

- Degenera su sistema *Master/Slave* per catene di più di 12 istanze.

- Lo sharding mantiene le copie di dati sulle singole catene *Primary/Secondaries*.

- Per realizzare un primo sharding servono almeno 8 istanze.

- La replica dei dati è uguale per tutti i database del sistema a meno di accorgimenti ad hoc su ogni macchina *Secondaries*.

**Riak**

<u>PRO</u>

- Distribuito e decentralizzato
- Ogni macchina è uguale alle altre.
- Ogni macchina può ricevere letture e scritture.
- Replica configurabile per ogni bucket per quanto concerne il numero di nodi che mantengono i dati.
- Da usare se si desidera scalare orizzontalmente e se accesso ai dati è prioritario via chiave.

<u>CONTRO</u>

- Lento.
- Queries di accessi ai dati limitate.

**Cassandra**

<u>PRO</u>

- Performante.
- A regime i nodi sono tutti uguali.
- Ogni macchina può ricevere letture e scritture.
- Da usare per scalare orizzontalmente e se serve ridimensionare/ampliare velocemente il cluster.

<u>CONTRO</u>

- I nodi inizialmente non sono tutti uguali, alcuni servono per contattare gli altri e costruire il cluster, va tenuto in considerazione in fase di start del cluster.
- Schema da definire.
- Column-Families da definire.

**Neo4j**

<u>PRO</u>

- Sistema Master/Slave
- Definisce dati e relazioni entrambi come documenti.
- Ogni macchina può ricevere letture e scritture.
- Permette join tra dati grazie alle relazione esplicitate e permette di evitare Map-Reduce.
- L'accesso ai dati avviene tramite linguaggi basati su query quali *Gremlin* e *Cypher;* quest'ultimo garantisce la transazionalità delle operazioni e le proprietà acide delle stesse.

<u>CONTRO</u>

- Non permette sharding.
- Lento.

# 8. Glossary

**Bitcask**
E' una applicazione Erlang che fornisce API per memorizzare o recuperare informazioni sotto forma di Key-Value in una hash-table LSF (Log-Structured File-system) che permette scritture ed accessi molto rapidi.

**CRUD**
Create, Read, Update, Delete : sono le operazioni di base di sistemi Key-Value.

**Data Model**
E' un modello astratto che permette di documentare ed organizzare, tramite testo e simboli, i dati o meglio l'informazione. [Hoberman]

**Erlang**
Linguaggio nato per ottimizzare disponibilità dell'informazione, concorrenza e tolleranza ai gusti.

**Gossip**
Protocollo di comunicazione peer-to-peer per scoprire e condividere informazioni di varia natura sui vari nodi che compongono il cluster. Le informazioni sono subito disponibili localmente su ogni nodo in modo da poter essere usate immediatamente al proprio restart.[12]

**HDFS**
Distributed File-System used by Hbase.

**Horizontal Partitioning - Sharding**
Memorizzazione di record di informazione su diverse macchine in base ad una data chiave.

**Horizontal Scaling – Scale out (Scalabilità Orizzontale)**
La capacità di distribuire su più macchine sia i dati che il carico delle operazioni per accedere, modificare e memorizzare gli stessi. Tutto questo senza che le macchine condividano ne' RAM ne' disco.

**JSON (Java Script Object Notation)**
Formato adatto allo scambio dati client-server usato molto per la semplicità della propria rappresentazione.

**MVCC (Multi-Versioning Cuncurrency Control)**
bla bla bla bla....

**OLTP**
It's a metodology to provide end user to access large amounts of data in rapid manner. Systems that facilitate and manage transaction-oriented applications. Is's used to refer to OLTP as the processing in which the system responds immediately to user requests, like Automatic Teller Machine of a Bank (ATM).

**OPLOG [MongoDB]**
Abbreviazione di OPeration LOG, è uno spazio usato per mantenere le informazioni delle operazioni che hanno modificato i dati memorizzati nel database. Il file viene scritto da una macchina Primary e messo a disposizione delle macchine Secondaries che, leggendolo, aggiornano in modo asincrono le modifiche che si sono attuate sul Primary. Si può vedere come lo strumento in grado di mantenere le repliche eventualmente consistenti nel tempo.

**Partizionamento Verticale**

Parte di un singolo record viene memorizzato su server diversi.

**Paxos**

Definizione di Paxos.

**Scalabilità Orizzontale**

Possiamo vederla come la possibilità di migliorare le performance di un sistema aumentando il numero delle macchine che esso ha a disposizione. L'aggiunta continua di machine permette di avere le stesse non troppo costose e di facile reperibilità sul mercato.

**Scalabilità Verticale (Vertical Scaling, Scale up)**

Ovvero il potenziamento del sistema dato dal potenziamento della macchina atto a mantenerlo: vengono quindi aumentate il numero di CPU oltre che RAM e disco. Chiaramente questa scalabilità, tipica dei RDBMS fa lievitare i costi dell'hardware.

**Thrift**

E' un framework dotato di un motore per la generazione di codice al fine di costrire servizi per i maggiori linguaggi di programmazione[18].

**Yum - Yellow dog Updater Modified**

E' il sistema di aggiornamento ufficiale per Fedora, affianca *rpm*, si appoggia a repository esterni per scaricare o aggiornare software sul sistema. La caratteristica principale di *yum* è quella di risolvere autonomamente le dipendenze necessarie per far si che i software da scaricare funzionino correttamente.